\documentclass[twocolumn]{article}
\usepackage[margin=0.6in]{geometry}

\usepackage[T1]{fontenc}
\usepackage{tikz}
\usetikzlibrary{backgrounds,patterns}
\usepackage{amsmath}
\usepackage{amsfonts}
\usepackage{hyperref}
\usepackage{filecontents}
\usepackage{xspace}
\usepackage{cleveref}
\usepackage{listings}
\usepackage{booktabs}
\usepackage{multirow}
\usepackage{tabularx}
\usepackage[table]{xcolor}
\usepackage{wasysym}
\usepackage{url}
\usepackage{subcaption}
\usepackage{minted}
\usepackage{pifont}
\usepackage{bm}
\usepackage{adjustbox}
\usepackage{enumitem}
\usepackage{inconsolata}

\usepackage{mathpartir}

\usepackage{nicematrix}
\usepackage{pgfmath}

\usepackage{fontawesome5}
\usepackage{caption}
\usepackage{longtable}

\usepackage{amsthm}
\theoremstyle{definition}

\setminted{
  linenos=false,
  numbersep=8pt,
  frame=none,
  xleftmargin=3pt,
  style=friendly,
  fontsize=\small,
  breaklines=true
}

\newcommand\lt[1]{{\lstinline!#1!}}

\crefname{section}{§}{§§} 
\crefname{subsection}{§}{§§} 
\crefname{chapter}{§}{§§} 

\newcommand{\work}[1]{\textsc{#1}}
\newcommand{\tool}[0]{\work{STITCH}\xspace}

\newcommand{\customsection}[1]{%
  \vspace{0.5em}
  \noindent\textbf{#1.}\xspace
}

\newcommand{\customsectionnoperiod}[1]{%
  \vspace{0.5em}
  \noindent\textbf{#1}\xspace
}

\definecolor{dkred}{RGB}{150, 0, 0}
\definecolor{tblue}{RGB}{0, 90, 181}

\newcommand{\circled}[1]{%
  \tikz[baseline=(char.base)]{
    \node[shape=circle,draw,inner sep=1pt,fill=black] (char) {\textcolor{white}{#1}};
  }%
}

\newcolumntype{R}[2]{%
    >{\adjustbox{angle=#1,lap=\width-(#2)}\bgroup}%
    l%
    <{\egroup}%
}

\newcolumntype{B}[1]{>{\raggedleft\arraybackslash}p{#1}}

\usepackage{tcolorbox}
\tcbuselibrary{listings,skins}
\usepackage{xparse}

\definecolor{inputcolor}{RGB}{59, 130, 246}
\definecolor{outputcolor}{RGB}{34, 197, 94}
\definecolor{inoutcolor}{RGB}{147, 51, 234}
\definecolor{codebg}{RGB}{253, 253, 254}
\definecolor{codeframe}{RGB}{226, 232, 240}
\definecolor{tabbg}{RGB}{226, 232, 240}
\definecolor{codeblockaccent}{RGB}{99, 102, 241}

\newtcblisting{codeblock}[3][]{
  enhanced,
  colback=codebg,
  colframe=codeframe,
  boxrule=1pt,
  arc=3pt,
  borderline west={3pt}{0pt}{codeblockaccent},
  boxsep=0pt,
  left=9pt,
  right=6pt,
  top=0pt,
  bottom=0pt,
  before skip=24pt,
  after skip=24pt,
  listing only,
  listing options={
    language=C++,
    basicstyle=\ttfamily\small,
    keywordstyle=\color{blue!70!black}\bfseries,
    commentstyle=\color{green!50!black}\itshape,
    stringstyle=\color{orange!70!black},
    numbers=none,
    breaklines=true,
    tabsize=2,
    showstringspaces=false,
    columns=flexible,
    escapeinside={(*@}{@*)},
    texcl=true,
    emph={FUZZ_ASSERT,FUZZ_PARAM,FUZZ_PARAM_STR,FUZZ_PARAM_FILE,FUZZ_BAIL,FUZZ_SET_ATTR_INT,FUZZ_GET_ATTR_INT,FUZZ_SET_ATTR_STR,FUZZ_GET_ATTR_STR,
    FUZZ_SET_ATTR_PTR,FUZZ_GET_ATTR_PTR,FUZZ_SET_ATTR_INT_GLOBAL,FUZZ_GET_ATTR_INT_GLOBAL,FUZZ_SET_ATTR_STR_GLOBAL,FUZZ_GET_ATTR_STR_GLOBAL},
    emphstyle=\color{magenta!80!black}\bfseries
  },
  overlay={
    \ifx\relax#1\relax\else
      \node[anchor=south west, inner sep=0pt, font=\textsf\small] 
        at ([yshift=1pt,xshift=4pt]frame.north west) {\color{codeblockaccent}\texttt{\footnotesize #1}};
    \fi
    \ifx\relax#2\relax\else
      \node[anchor=south, inner sep=0pt] 
        at ([yshift=-0.5pt]frame.north) {#2};
    \fi
    \ifx\relax#3\relax\else
      \node[anchor=north, inner sep=0pt] 
        at ([yshift=0.5pt]frame.south) {#3};
    \fi
  },
}
\newtcblisting{codebox}{
  enhanced,
  colback=codebg,
  colframe=codeframe,
  boxrule=1pt,
  arc=3pt,
  boxsep=0pt,
  left=6pt,
  right=6pt,
  top=0pt,
  bottom=0pt,
  box align=top,
  listing only,
  listing options={
    language=C++,
    basicstyle=\ttfamily\small,
    keywordstyle=\color{blue!70!black}\bfseries,
    commentstyle=\color{green!50!black}\itshape,
    stringstyle=\color{orange!70!black},
    numbers=none,
    breaklines=true,
    tabsize=2,
    showstringspaces=false,
    columns=flexible,
    escapeinside={(*@}{@*)},
    texcl=true,
    emph={FUZZ_ASSERT,FUZZ_PARAM,FUZZ_PARAM_STR,FUZZ_PARAM_FILE,FUZZ_BAIL,FUZZ_SET_ATTR_INT,FUZZ_GET_ATTR_INT,FUZZ_SET_ATTR_STRING,FUZZ_GET_ATTR_STRING},
    emphstyle=\color{magenta!80!black}\bfseries
  },
}
\newtcolorbox{mathcodeblock}[2]{
  enhanced,
  colback=codebg,
  colframe=codeframe,
  boxrule=0.5pt,
  arc=3pt,
  boxsep=4pt,
  left=6pt,
  right=6pt,
  top=2pt,
  bottom=2pt,
  before skip=16pt,
  after skip=16pt,
  overlay={
    \ifx\relax#1\relax\else
      \node[anchor=south, inner sep=0pt] 
        at ([yshift=-0.5pt]frame.north) {#1};
    \fi
    \ifx\relax#2\relax\else
      \node[anchor=north, inner sep=0pt] 
        at ([yshift=0.5pt]frame.south) {#2};
    \fi
  },
}

\begin{document}

\title{\bf Automatic, Expressive, and Scalable Fuzzing with Stitching}

\author{
  Harrison Green\\
  \texttt{harrisog@cmu.edu}\\
  Carnegie Mellon University
  \and
  Fraser Brown\\
  \texttt{fraserb@cmu.edu}\\
  Carnegie Mellon University
  \and
  Claire Le Goues\\
  \texttt{clegoues@cmu.edu}\\
  Carnegie Mellon University
}

\date{}

\maketitle

\newcommand{\numpatched}[0]{73\xspace}

\begin{abstract}
Fuzzing is a powerful technique for finding bugs in software libraries, but
scaling it remains difficult. Automated harness generation commits to fixed API
sequences at synthesis time, limiting the behaviors each harness can test.
Approaches that instead explore new sequences dynamically lack the expressiveness
to model real-world usage constraints 
leading to false positives from
straightforward API misuse.

We propose \emph{stitching}, a technique that encodes API usage constraints in
pieces that a fuzzer dynamically assembles at runtime. 
A static type system governs how objects flow between blocks, while a
dynamically-checked extrinsic typestate tracks arbitrary metadata across
blocks, enabling specifications to express rich semantic constraints such as
object state dependencies and cross-function preconditions. This allows a single
specification to describe an open-ended space of valid API interactions that the
fuzzer explores guided by coverage feedback.

We implement stitching in \tool, using LLMs to automatically configure projects for fuzzing, synthesize a specification, triage crashes, and repair the specification itself. We evaluated \tool against four state-of-the-art tools on 33 benchmarks, where it achieved the highest code coverage on 21 and found 30 true-positive bugs compared to 10 by all other tools combined, with substantially higher precision (70\% vs.\ 12\% for the next-best LLM-based tool). Deployed automatically on 1365 widely used open-source projects, \tool discovered 131 new bugs across 102 projects, \numpatched of which have already been patched.
\end{abstract}

\section{Introduction}

Dynamic software testing is a cornerstone of modern
software reliability and security.
Unit and integration tests are fundamental
parts of the software engineering process, and
fuzzing---automated random testing---is the dominant
approach for automatically finding security vulnerabilities. 
Fuzzers deployed as part of \work{oss-fuzz}, for example, have already discovered more than 50,000 bugs across more than 1,000 open source projects~\cite{oss-fuzz}.

Since the early 2000s, researchers have focused on automation to scale testing with the size and complexity of modern
software. 
This automation occurs along two complementary axes:
synthesizing testing artifacts themselves, 
and improving how existing testing engines explore program behavior.
The latter includes pioneering work like 
\work{QuickCheck}~\cite{quickcheck} and \work{Randoop}~\cite{randoop}, to more recent work in coverage-guided fuzzing~\cite{afl,afl++,libafl,honggfuzz,syzkaller,fuzzilli2} and property-based testing~\cite{jqf,hypothesis,GoogleFuzzTest,vikram2023can};
the former includes work that synthesizes test suites~\cite{evosuite,catlm} and fuzz harnesses~\cite{fudge, fuzzgen, utopia,promptfuzz, promefuzz,oss-fuzz-gen,ogharn}.

Unfortunately, these testing approaches don't scale to completely
exercising most programs' behaviors.
As software grows linearly in compositional freedom---becoming richer
through new functions and objects---the space of possible interactions
explodes combinatorially---four new functions can now be invoked in
sixty-four new ways.
Unit testing every behavior is intractable.
Automated approaches like property-based testing and fuzzing
still don't cover the massive space of higher-order interactions,
since they're limited to invoking specific, finite API sequences.
Even if it \emph{were} possible to invoke all possible API calls
in all possible orders over all possible data, we would only
celebrate for so long: most library API interactions are simply invalid.
Consumers of a library API must follow specific usage constraints
from the obvious and structural (e.g., a function only accepts the right
data types) to the higher-level and semantic (e.g., an object must
be in a specific state before a certain function may use it). 
Unfortunately, these usage constraints are often difficult both to interpret and enforce.
Testing systems that do not model them spend almost all of their time discovering (expected) API misuses
instead of real bugs.
In sum, the key challenge for automated testing is how to scale to
exercise \emph{all} the \emph{valid} interactions in a
program---and nothing more. 

In this paper, we propose \emph{stitching}, a testing technique that encodes
API usage constraints in composable ``blocks'' that a fuzzer
automatically ``stitches'' together at runtime.
Each block captures a set of valid invocations of a single
library function; together, blocks describe valid function sequences.
Stitching allows a fuzzer to
explore the combinatorially-large interaction space of functions
and objects while avoiding invalid parts of that space.

There are two key contributions that make \emph{stitching}
work well in practice.
First, we recognize that real-world API constraints
often go beyond what a simple static type system
can capture. 
Thus, we develop a hybrid approach for expressing
constraints:
a static type system that governs how objects flow between code blocks coupled with extrinsic typestate that tracks arbitrary,
dynamically-updated metadata about those objects across block executions.
Dynamic typestate makes it possible to encode
rich, semantic constraints that are enforced at runtime:
specifications read and write metadata on objects
(e.g., whether a file handle is open),
and then elegantly ``bail'' when constraints over
that metadata are violated (e.g., by attempting to read
from a closed file).
As a result, the fuzzing search process is both the mechanism that finds new coverage in the target, but also the mechanism that identifies new valid sequences of library calls (those which execute without bailing).

Second, API usage constraints exist as source
code, comments, documentation, error messages, and more. 
LLMs are well-suited to
retrieving and synthesizing these disparate sources of information into (correct
enough) constraints.
Starting from this observation, we designed \tool, an LLM-driven
system that drafts initial specifications and iteratively
refines them based on compilation feedback and observed test outcomes.
We deployed \tool on 1365 widely used open-source projects, discovering 131 new bugs across 102 projects
\emph{without any manual intervention}---the system
configures projects with fuzzer and sanitizer instrumentation, infers
specifications, tests code, refines specifications, and minimizes
and triages crashes found during testing. 

\customsection{Contributions}
We make the following contributions:
\begin{itemize}
  \item \textbf{\emph{Stitching}}: A new technique for encoding arbitrary API constraints in a fuzzable specification.
  \item \textbf{\tool}: An implementation of \emph{stitching} for C/C++ code capable of testing projects completely independently.
  \item A coverage evaluation comparing \tool to state-of-the-art systems: \tool achieves higher coverage and finds more bugs (with more precision) than other SOTA systems, while being faster and more cost-effective (in terms of \$/bug).
  \item A real-world evaluation: We applied \tool to 1365 open source projects entirely automatically and reported 131 bugs in 102 projects, \numpatched of which have already been patched.
\end{itemize}

\section{Motivation and Challenges}

To motivate our approach, we present a case study of a bug
\tool discovered that had eluded prior testing efforts.
Listing \ref{lst:pupnp-crash} shows a minimized reproducer for a crash \tool discovered in \texttt{pupnp}, a popular SDK for building UPnP (Universal Plug and Play) devices.
The project is a dependency of widely used software like VLC Media Player,
and has been fuzzed extensively through \work{OSS-Fuzz}. However, its test
suites (previously) had no
tests involving elements and attributes. 
\tool automatically discovered this bug, which crashes the XML handling
subsystem by adding a namespaced attribute with the same name as a
non-namespaced attribute; it also produced a minimized external reproducer and complete bug report.
After manual verification, we submitted the issue to the maintainers,
who responded, ``Amazing work, congratulations!''
The bug revealed a broader class of namespace-related issues, all patched within 24 hours of disclosure.

\begin{figure}
  \captionsetup{name=Listing}
\begin{codebox}
#include "upnp/ixml.h"
int main() {
  Document *doc;
  Element *el;
  Attr *a, *ans, *repl, *repl2;

  // 1) Create a new document and element
  doc = createDocument();
  el = createElement(doc, "E");

  // 2) Attach a non-ns attribute
  a = createAttr(doc, "x");
  if (setAttrNode(el, a, &repl) != SUCCESS)
    return 0;

  // 3) Attach a ns attribute with the same name
  ans = createAttrNS(doc, "urn:ns", "p:x");
  setAttrNodeNS(el, ans, &repl2); // Crash!
  return 0;
}
\end{codebox}
\caption{A testcase triggering a crash in \texttt{pupnp}. Function and type names have been shortened and error handling has been removed for brevity.}
\label{lst:pupnp-crash}
\end{figure}

Although the testcase is simple, the bug is difficult to discover
automatically:

\customsection{Challenge 1: Combinatorial search space}
This bug requires six API calls to occur in order: 
create a document and element, construct two attributes with the same name (one namespaced, one not), and assign them to the element.
Meanwhile, the space of possible API sequences in general is combinatorially large.  For
the XML subsystem alone (81 functions), naively sampling function 6-tuples
yields a
$\frac{1}{81^6}$ (roughly 1 in 282 billion) chance of hitting the right
sequence. 
This effectively dooms any approach that commits to fixed API
sequences before fuzzing. The XML parser's existing manual harness
(\texttt{FuzzIxml}), has been integrated in \work{OSS-Fuzz} for three years, but
does not invoke these attribute functions.
Automated harness generation~\cite{fuzzgen,fudge,utopia,ogharn,promptfuzz,promefuzz,oss-fuzz-gen} broadens fuzz coverage by synthesizing harnesses 
for different API sequences.
However, each harness still commits
to a fixed API sequence at synthesis time.
To illustrate, we
ran \work{PromeFuzz} on the XML subsystem (81 functions).
It produced 31
harnesses totaling \textasciitilde 4000 lines of code. Each implicated function was invoked by at least one harness, but no harness
invoked all six.

\customsection{Challenge 2: Complex API usage constraints}
Recent work allows fuzzers to adaptively explore new API sequences at
runtime~\cite{hopper,graphfuzz}, addressing the combinatorial problem but
encountering a new one: most API sequences are invalid or uninteresting.
API
consumers must follow strict but informally documented usage constraints, both
structural (e.g., providing the right types to each function) and semantic
(e.g., 
validating return values, ensuring attributes created with
\texttt{createAttr} are only passed to \texttt{setAttrNode} rather than
the namespaced version (\texttt{setAttrNodeNS}) and vice-versa, and ensuring the element and attribute
provided to \texttt{setAttrNode} share the same parent document).
A fuzzer that ignores these constraints wastes time finding crashes caused by
straightforward and uninteresting API misuse. For example, 
\work{Hopper}~\cite{hopper} randomly searches through sequences heuristically, and
mostly finds false
positive crashes on this target. \work{GraphFuzz}~\cite{graphfuzz} follows a manual specification
of what objects functions consume and produce. However, its specification language can only express
basic object lifetime constraints, and cannot (for example) model the constraint
that elements and attributes passed to \texttt{setAttrNode} must have the same
parent document.

\customsection{Our Solution: Stitching}
\emph{Stitching} addresses both challenges through dynamic sequence generation
guided by a rich specification language.  
The specification consists of small code
blocks (like \work{GraphFuzz}~\cite{graphfuzz}) that describe how to invoke single API functions.  
At runtime, the fuzzing engine stitches these blocks together, guided by
coverage feedback, to dynamically explore testcases. 
The fuzzer can thus explore 
millions of possible API sequences, rather than
the few hundred an auto-harnesser can cover. To capture
complex usage constraints, we pair a static type system (ensuring structural
correctness) with 
a dynamically-checked extrinsic typestate: code blocks can
read and write arbitrary metadata on the objects they interact
with, and this metadata persists across blocks. This means specifications can
track and enforce rich semantic properties, such as whether an attribute is
namespaced or two objects share the same parent document. Blocks that encounter
violated constraints simply \emph{bail} (exit cleanly), focusing the search on
valid API usage. Because specifications are expressed as valid source code with
lightweight macros, they are also well-suited for LLM-based synthesis.

\section{Stitching}
\label{sec:stitching}

With stitching, the testable surface of an API is encoded in a fuzzable specification.
 At runtime, the engine samples from this specification to dynamically construct
 and execute testcases. 
 In this section, we formalize the core specification format and execution semantics as it applies to any target language (\S\ref{sec:formal-semantics}), discuss specific implementation details for C/C++ (\S\ref{sec:stitching-c-cpp}), and describe what these code blocks look like in practice when modeling real-world APIs (\S\ref{sec:modeling-real-world-apis}).

\subsection{Formal Semantics}
\label{sec:formal-semantics}

Here we present a formalization of \emph{stitching} using a big-step operational semantics. We describe the structure of code blocks, how they appear in testcases as block instances, and what it means to execute a testcase.

\customsection{Types and Objects}
Let $\mathcal{T}$ be a finite set of types and $\mathcal{O}$ a universe of object identifiers, with $\tau(o) \in \mathcal{T}$ denoting the type of object $o$.

\customsection{Code Blocks}
A \emph{code block} $b$ is a 4-tuple:
$$b = (\mathrm{Code}_b, \mathrm{In}_b, \mathrm{Out}_b, \mathcal{F}_b)$$
consisting of some executable native code $\mathrm{Code}_b$ along with a well-defined interface of $n$ typed inputs $\mathrm{In}_b \in \mathcal{T}^n$, $m$ typed outputs $\mathrm{Out}_b \in \mathcal{T}^m$, and a space of fuzzable parameters $\mathcal{F}_b$ which is defined implicitly in our implementation (see \ref{sec:codeblocks}).

\customsection{Block Instance}
A \emph{block instance} $(b, \vec{r}, f)$ describes a single instance of code block (as encapsulated in a testcase), where:
\begin{itemize}[leftmargin=2em, noitemsep, topsep=0pt]
\item $b$ is a code block,
\item $\vec{r} \in \mathbb{N}^{|\text{In}_b|}$ is a sequence of \emph{reference indices}, mapping each input type to a same-typed output from an earlier block instance, and
\item $f \in \mathcal{F}_b$ contains concrete values for the fuzzable parameters
\end{itemize}

\customsection{Testcases}
A \emph{testcase} $T$ is a sequence of block instances:
$$
T = [(b_0, \vec{r}_0, f_0), \ldots, (b_{\ell-1}, \vec{r}_{\ell-1}, f_{\ell-1})]
$$

The reference indices in the testcase implicitly form a directed graph, mapping outputs of one block to inputs of another block. A testcase is well formed if this graph is acyclic, each output is referenced at most once, and block instances are ordered topologically. In other words, since a testcase invokes block instances in order, all the inputs for block instance $t$ must have been produced \emph{before} index $t$.

Formally, a testcase $T$ is \emph{well-formed} iff: 

\begin{itemize}[leftmargin=2em, noitemsep, topsep=0pt]
\item (\textbf{Backward References}) References must point to outputs from earlier instances. Let $N_i = \sum_{j=0}^{i-1} |\text{Out}_{b_j}|$ be the total number of objects produced before invocation $i$. Then every reference index must point to a previous output object: $\forall k \in \vec{r}_i.\; k < N_i$.
\item (\textbf{Single-Use Outputs}) Every reference must point to a distinct object (outputs cannot be used more than once): $\forall i,i',j,j'.\; (i \neq i' \lor j \neq j') \implies \vec{r}_i[j] \neq \vec{r}_{i'}[j']$.
\item (\textbf{Type Correctness}) For each instance $i$ with input types $\mathrm{In}_{b_i} = [\tau_0, \ldots, \tau_{n-1}]$, every reference must point to an object of the correct type: $\forall j < n.\; \tau(\text{output at index } \vec{r}_i[j]) = \tau_j$, where ``output at index $k$'' means: if $k = \sum_{j=0}^{p-1} |\text{Out}_{b_j}| + q$ for some $p,q$, then the $q$-th output of the $p$-th invocation.
\end{itemize}

\customsection{Extrinsic Typestate}
A key language design choice is the introduction of extrinsic typestate. We track both per-object ($\mathcal{M}_o$) and global ($\mathcal{M}_g$) typestate via arbitrary key-value pairs:
\begin{align*}
\mathcal{M}_o&: \mathcal{O} \rightarrow (\mathcal{K} \rightharpoonup \mathcal{V}) \\
\mathcal{M}_g&: \mathcal{K} \rightharpoonup \mathcal{V}
\end{align*}

\customsection{Block Execution}
Executing a native code block $\mathrm{Code}_b$ requires suitable input objects $\vec{o}$, fuzzable parameters $f$, and the typestate stores $\mathcal{M}_o$ and $\mathcal{M}_g$. Execution produces either a set of output objects $\vec{o}'$ and updated typestate stores $\mathcal{M}_o'$ and $\mathcal{M}_g'$ or a \emph{bailout} signal $\mathsf{Bail}$ indicating a non-recoverable failure.
$$
[\![\mathrm{Code}_b]\!] : \mathcal{O}^n \times \mathcal{F}_b \times \mathcal{M}_o \times \mathcal{M}_g \to (\mathcal{O}^m \times \mathcal{M}_o' \times \mathcal{M}_g') \uplus \mathsf{Bail}
$$

\customsection{Runtime State}
The runtime state $\Sigma = (\vec{O}, \mathcal{M}_o, \mathcal{M}_g)$ tracks $\vec{O}$ the flat sequence of all objects produced, and the extrinsic typestates stores $\mathcal{M}_o$ and $\mathcal{M}_g$.

\customsection{Testcase Execution}
Executing a testcase $T$ involves starting from the initial runtime state $\Sigma_0 = ([], \emptyset, \emptyset)$ and executing the block instances in $T$ in sequence, updating the runtime state after each block instance, or capturing a bailout signal at a particular block instance index $\mathsf{Bail}@i$:
$$
\Sigma, T \Longrightarrow^* \Sigma' \uplus \mathsf{Bail}@i
$$

\customsection{Operational Semantics}
\Cref{fig:semantics} presents the operational semantics. Part~(a) defines single-step execution with judgment $\Sigma, (b, \vec{r}, f) \Longrightarrow \Sigma' \uplus \mathsf{Bail}$: successful execution consumes inputs, produces typed outputs, and updates the typestate stores; blocks may also bail (e.g., when dynamic constraints fail). Part~(b) lifts this to multi-step testcase execution over sequences of block instances.

\customsection{Crashes}
The ultimate goal of stitching is to find testcases which trigger bugs, segmentation faults, assertions, sanitizer exceptions, or other non-recoverable failures. We intentionally do not model these cases in our semantics, but rather treat them as an external property of testcase execution.

\begin{figure*}[t]
  \centering
  
  \setlength{\fboxrule}{0.6pt}
  \setlength{\fboxsep}{0pt}
  
  \fcolorbox{black!35}{white}{%
    \begin{minipage}{\linewidth}
      \vspace{0pt}
      {
        \setlength{\fboxsep}{1em}
        \colorbox{white}{%
          \begin{minipage}{0.965\linewidth}
            \paragraph{(a) Single-Step Block Execution: $\Sigma, (b, \vec{r}, f) \Longrightarrow \Sigma' \uplus \mathsf{Bail}$}
            \begin{mathpar}
            \inferrule[Exec-Success]{
              \vec{o} = \mathit{resolve}(\vec{r}, \vec{O}) \\
              [\![\mathrm{Code}_b]\!](\vec{o}, f, \mathcal{M}_o, \mathcal{M}_g) \Downarrow (\vec{o}', \mathcal{M}_o', \mathcal{M}_g') \\
              |\vec{o}'| = |\mathrm{Out}_b| \land \forall i.\; \tau(o'_i) = \mathrm{Out}_b[i]
            }{
              (\vec{O}, \mathcal{M}_o, \mathcal{M}_g), (b, \vec{r}, f) \Longrightarrow (\vec{O} \mathbin{+\!\!+} [\vec{o}'], \mathcal{M}_o', \mathcal{M}_g')
            }
            
            \inferrule[Exec-Bail]{
              \vec{o} = \mathit{resolve}(\vec{r}, \vec{O}) \\
              [\![\mathrm{Code}_b]\!](\vec{o}, f, \mathcal{M}_o, \mathcal{M}_g) \Downarrow \mathsf{Bail}
            }{
              (\vec{O}, \mathcal{M}_o, \mathcal{M}_g), (b, \vec{r}, f) \Longrightarrow \mathsf{Bail}
            }
            \end{mathpar}
          \end{minipage}%
        }
      }
      \par\nointerlineskip
      {
        \setlength{\fboxsep}{1em}
        \colorbox{black!6}{%
          \begin{minipage}{0.965\linewidth}
            \paragraph{(b) Multi-Step Testcase Execution: $\Sigma, T \Longrightarrow^* \Sigma' \uplus \mathsf{Bail}@i$}
            \begin{mathpar}
            \inferrule[Seq-Empty]{ }{
              \Sigma, [] \Longrightarrow^* \Sigma
            }
            
            \inferrule[Seq-Step]{
              \Sigma, (b, \vec{r}, f) \Longrightarrow \Sigma' \\
              \Sigma', T \Longrightarrow^* \Sigma''
            }{
              \Sigma, [(b, \vec{r}, f) \mid T] \Longrightarrow^* \Sigma''
            }
            
            \inferrule[Seq-Bail]{
              \Sigma, (b, \vec{r}, f) \Longrightarrow \mathsf{Bail}
            }{
              \Sigma, [(b, \vec{r}, f) \mid T] \Longrightarrow^* \mathsf{Bail}@0
            }
            
            \inferrule[Seq-Bail-Later]{
              \Sigma, (b, \vec{r}, f) \Longrightarrow \Sigma' \\
              \Sigma', T \Longrightarrow^* \mathsf{Bail}@i
            }{
              \Sigma, [(b, \vec{r}, f) \mid T] \Longrightarrow^* \mathsf{Bail}@(i+1)
            }
            \end{mathpar}
          \end{minipage}%
        }
      }
  
      \vspace{0pt}
    \end{minipage}
  }
  
  \caption{Operational semantics of block and testcase execution. Resolution is defined as: $\mathit{resolve}(\vec{r}, \vec{O}) = [\vec{O}[k] \mid k \in \vec{r}]$}
  \label{fig:semantics}
\end{figure*}

\subsection{Stitching for C/C++}
\label{sec:stitching-c-cpp}

In our implementation, this \emph{fuzzable specification} is encoded as a JSON file containing code blocks, each with a \texttt{code} string (C++ code) and lists of inputs and outputs. We describe code blocks in \S\ref{sec:codeblocks} and the fuzzing engine in \S\ref{sec:engine}. 

\subsubsection{Code Blocks in C/C++}
\label{sec:codeblocks}

Each code block can contain valid function-body code (statements, local declarations, expressions, etc.) but cannot contain function declarations or definitions.
During compilation, the contents of each code block is inserted (verbatim except for syntactic sugar replacements) into the body of a generated driver function in an amalgamated \texttt{harness.cpp} file. The first code block becomes \texttt{func\_0}, the second becomes \texttt{func\_1}, etc. This file is then compiled and linked against the target library.

Input and output types ($\mathcal{T}$) correspond to actual C/C++ types.
Typically these are C/C++ structs or classes, and occasionally primitive types like \texttt{int}s when they carry API-level meaning (e.g., as integer file handles or resource identifiers).
Objects in $\mathcal{O}$ correspond to pointers to these types.

\customsection{Bailout}
Code blocks indicate acceptable (non-crashing) failures with the \texttt{FUZZ\_BAIL()} macro:
\begin{codebox}
noreturn void FUZZ_BAIL() // $\to \mathsf{Bail}$
\end{codebox}

\customsection{Fuzzable Parameters}
The fuzzable parameter space $\mathcal{F}_b$ is defined implicitly by the use of \texttt{FUZZ\_PARAM} macros inside the code block. A code block can invoke these macros to request random (fuzzable) data of a specified type, similar to existing conventions like \texttt{FuzzedDataProvider} in \texttt{libFuzzer}~\cite{libfuzzer}.~\footnote{Unlike \texttt{FuzzedDataProvider}, however, these values are not deserialized ad-hoc from a global fuzzer buffer, but rather are treated as first-class citizens and (by default) preserved during mutation, not susceptible to the \emph{havoc effect}~\cite{li2024havoc}.}

\begin{codebox}
(*@\hbox to 15em{T*~~~~~~~~~~\color{magenta!80!black}{\bfseries FUZZ\_PARAM}\color{black}(T)\hss}@*) // random T
(*@\hbox to 15em{std::string \color{magenta!80!black}{\bfseries FUZZ\_PARAM\_STR}\color{black}()\hss}@*) // random string
(*@\hbox to 15em{std::string \color{magenta!80!black}{\bfseries FUZZ\_PARAM\_FILE}\color{black}()\hss}@*) // random file data
\end{codebox}

\customsection{Extrinsic Typestate Interface}
Code blocks observe and update extrinsic typestate (or ``metadata'')
on objects ($\mathcal{M}_o$) and the global state ($\mathcal{M}_g$)
through a set of macros
($\texttt{T} \in \{\texttt{INT}, \texttt{STR}, \texttt{PTR}\}$)
exposed by the runtime.

\begin{codebox}
(*@\hbox to 16em{\color{blue!70!black}\bfseries void \color{magenta!80!black}FUZZ\_SET\_ATTR\_T\color{black}\normalfont($o$, $k$, $v$)\hss}@*)// $M_o[o][k \mapsto v]$
(*@\hbox to 16em{T~~~~\color{magenta!80!black}\bfseries FUZZ\_GET\_ATTR\_T\color{black}\normalfont($o$, $k$)\hss}@*)// $!M_o[o][k]$
(*@\hbox to 16em{\color{blue!70!black}\bfseries void \color{magenta!80!black}FUZZ\_SET\_ATTR\_T\_GLOBAL\color{black}\normalfont($k$, $v$)\hss}@*)// $M_g[k \mapsto v]$
(*@\hbox to 16em{T~~~~\color{magenta!80!black}\bfseries FUZZ\_GET\_ATTR\_T\_GLOBAL\color{black}\normalfont($k$)\hss}@*)// $!M_g[k]$
\end{codebox}

\subsubsection{Fuzzing Engine}
\label{sec:engine}

Given a specification of code blocks, the role of the fuzzing engine is to construct new testcases repeatedly (by stitching together code blocks) and invoke them, with the objective of maximizing coverage of the target API.
Unlike a traditional fuzzer which maintains a corpus of byte-sequence \emph{inputs} (provided to a single fuzzing entrypoint), each corpus entry in our fuzzer is a fully-formed \emph{testcase} consisting of a sequence of stitched-together code blocks and their concrete fuzzable parameter values. Invoking a corpus entry invokes each code block in sequence, with fuzzable values injected, and inputs and outputs properly assigned.

The heart of a fuzzing engine generates and mutates well-formed testcases.
Ensuring that synthesized testcases execute without bailing is undecidable;
instead, we generate testcases randomly to discover those that execute
successfully, akin to \emph{rejection sampling} used by testing frameworks like
\work{QuickCheck}~\cite{quickcheck}.  That is, 
our mutation algorithms only guarantee well-formedness, and we rely on the coverage-guided search to discover \emph{successful} testcases.

\customsection{Mutation Algorithms}
We define two kinds of mutations on testcases: structural
mutations---conceptually, scrambling code blocks---and
parameter mutations---straightforwardly changing fuzzable
parameter values.

\customsection{Structural Mutations}
Structural mutations can change the makeup and order of block instances in the
testcase and the connectivity of objects ($\vec{r}$) between them. We define four
operators for a testcase $T$:

\begin{itemize}[leftmargin=2em, noitemsep, topsep=0pt]
  \item \textbf{Regenerate}: selects a random code block as a seed for a fresh testcase, and adds other code blocks as necessary to construct requisite inputs and satisfy \emph{well-formedness}.
  \item \textbf{Crossover}: given a corpus of testcases $\vec{T}$, selects another testcase $T' \in \vec{T}$ and stitches at least one block instance from $T'$ into the current testcase.
  \item \textbf{FrontierExtend}: given a testcase that bails ($\Sigma, T \Downarrow \textsf{Bail}@i$), selects the \emph{frontier} instance $i-1$ and stitches a new block instance to it, either consuming one of its outputs or producing one of its inputs (if the testcase is not bailing, picks a random instance $i$ instead).
  \item \textbf{FrontierTrimRepair}: given a testcase that bails ($\Sigma, T \Downarrow \textsf{Bail}@i$), selects the \emph{frontier} instance $i-1$ and trims the graph to this point (removing failing or skipped downstream instances) (if the testcase is not bailing, picks a random instance $i$ instead).
\end{itemize}

\customsection{Parameter Mutations}
Parameter mutation changes the fuzzable parameter values $f$ that are provided to the code blocks (as requested by the \texttt{FUZZ\_PARAM} macros). For fixed-size \texttt{FUZZ\_PARAM} invocations, it directly mutates the serialized bytes. For variable-size strings and file data, it mutates both the underlying byte sequence and the size parameter. Concretely, our implementation repurposes existing byte-sequence mutators from \work{LibAFL}~\cite{libafl} and applies them to the serialized bytes of the fuzzable parameters. Additionally, for string and file data, users can provide \emph{hints} (i.e., good initial values) in the schema to use a small percentage of the time as a starting point for the mutation (see \S\ref{sec:data-generator-synthesis}).

\customsection{Crash Minimization}
When the fuzzing engine discovers a crash, it automatically minimizes the testcase to the smallest possible input that still triggers the crash by iteratively mutating the testcase and checking if it still triggers the crash, until convergence.

\customsection{External Reproducers}
Our semantics for C/C++ code blocks also allow the fuzzing engine to convert fuzzer-internal testcases into external reproducers (independently compilable C++ files), by effectively inlining each of the code blocks into the \texttt{main} function of a new C++ file, baking in the input/output linkage and fuzzable parameter values.

\subsection{Modeling Real-World APIs}
\label{sec:modeling-real-world-apis}

\begin{figure}
  \centering
  \resizebox{\linewidth}{!}{
    \includegraphics{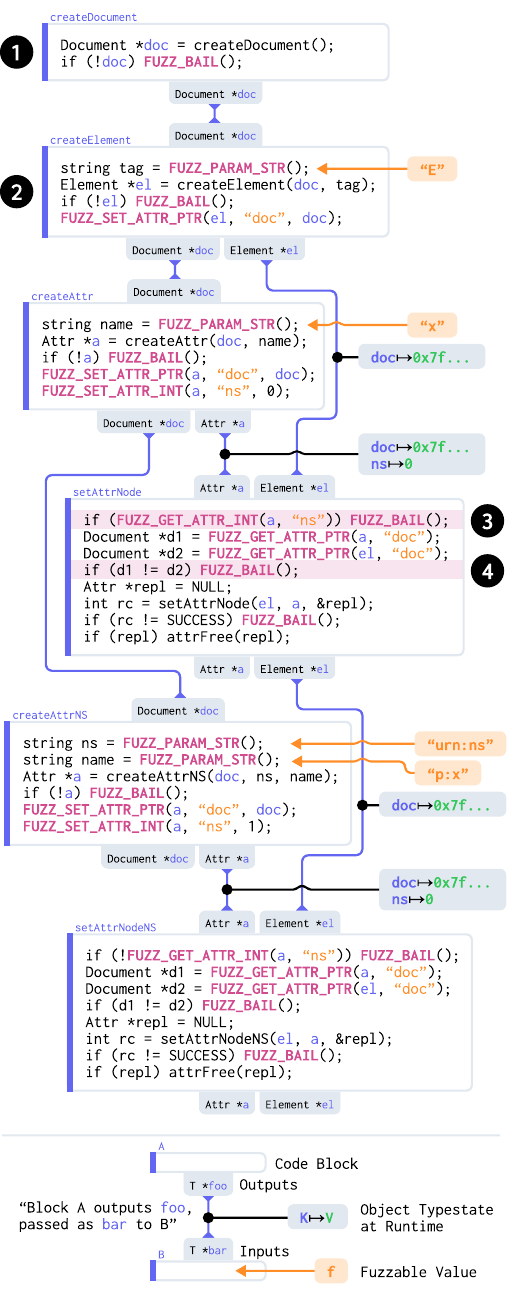}
  }
  \caption{Visualization of six code blocks \emph{stitched} together in a
  testcase, reproducing the same crash as \autoref{lst:pupnp-crash}.}
  \label{fig:example-graph}
\end{figure}

To put the pieces together,
in \autoref{fig:example-graph}, we show how six code blocks for functions in \texttt{pupnp} can be stitched together, hitting the same crash as in \autoref{lst:pupnp-crash}.
Note that the figure shows \emph{one} example of how these six code blocks can be stitched;
in practice, the engine will test millions of different combinations, changing the order, duplicating blocks, and combining these blocks with others from the library.

Looking at each code block separately, for a moment, we can observe how each function is individually specified. For example, the \texttt{createDocument} code block (\circled{1}) consists of C++ code that invokes the function \texttt{createDocument} and checks its return value (bailing if no document was produced). The newly created document is specified to be an output (indicated by the tab on the bottom of the block).

Next, in the \texttt{createElement} code block (\circled{2}), it receives a document as input (tab on the top), which in this case is connected to the prior code block, meaning that the engine will first call \texttt{createDocument} and then pass the resulting object to be used in the \texttt{createElement} block. Creating an element also takes a string tag, which we can request using \texttt{FUZZ\_PARAM\_STR}. The fuzzing engine can mutate this value, but in this testcase it has assigned the value \texttt{"E"}. In this block, we invoke \texttt{createElement}, check the return value, and then return both the document and newly created element as outputs for downstream blocks to use.

Unfortunately, naively combining code blocks based on their type signature will actually land us in trouble. We might accidentally pass an attribute created with \texttt{createAttr} to \texttt{setAttrNodeNS} (illegal!) or try to use an attribute and element created in different documents in the call to \texttt{setAttrNode} (illegal!).

Our solution is the use of extrinsic typestate. In a specification code blocks can set and get arbitrary metadata on objects and use it to implement dynamic validation checks. For example, we can enforce proper attribute correlation by setting the key \texttt{ns} to either \texttt{0} or \texttt{1} depending on which constructor it came from and then check this value in the calls that depend on it (\circled{3}). Similarly, when we create an attribute or element, we can store the pointer value of its parent document in the \texttt{doc} key, and then upon receiving inputs in the \texttt{setAttrNode} block, we can validate that these values match (\circled{4}). As a result, the engine is free to search for interesting combinations of APIs
without triggering a flood of false positives.

\begin{figure}
  \resizebox{\linewidth}{!}{
    \includegraphics{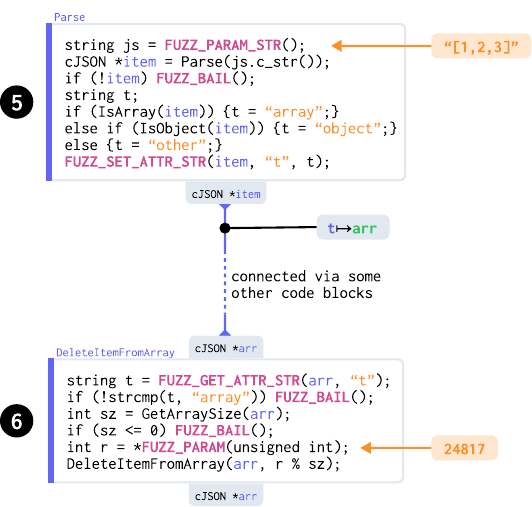}
  }
  \caption{Example of mixing extrinsic typestate with dynamic checkers.}
  \label{fig:cjson}
\end{figure}

The free-form nature of code blocks and constraint checks scales naturally to
more complex situations. Consider the \texttt{cJSON} library for parsing and
manipulating JSON. \autoref{fig:cjson} shows code blocks for two of its
functions: \texttt{Parse}, which attempts to parse a JSON string and returns the
resulting object, and \texttt{DeleteItemFromArray}, which removes an element at a
given index.

Even these two functions present several challenges. \texttt{cJSON} uses a single
pointer type \texttt{cJSON~*} to represent every internal variant (objects,
arrays, integers, strings, etc.) a pattern common in C/C++ code. Functions like
\texttt{DeleteItemFromArray} require that the pointer actually refers to an
array, but when calling \texttt{Parse} we cannot know the resulting type in
advance: it depends on the JSON string, which the fuzzer mutates.
\texttt{DeleteItemFromArray} also requires a valid index; passing 10 to a
three-element array is illegal.

Because code blocks are free-form code, we can enforce these constraints
dynamically. After parsing, we inspect the result with other functions from the library like \texttt{IsArray} and \texttt{IsObject} and set metadata accordingly for
downstream blocks (\circled{5}). In \texttt{DeleteItemFromArray}, we check this
metadata to validate the type, then call another function from the library \texttt{GetArraySize} to obtain the array
length, bail if it is empty, and cast a random fuzzable value into range
otherwise (\circled{6}).

\section{\tool}
\label{sec:design}

Our stitching implementation can be used as a standalone, manually-configured fuzzer. However, manual configuration is laborious, requiring deep understanding of the target library's usage constraints.
These constraints are often under-specified and scattered throughout the
codebase in natural language, intended for human readers who can fill in the
blanks. LLMs, trained extensively on both code and natural language, are
well-suited to synthesize this information into stitchable specifications.

That said, specification errors are inevitable, even for human experts.
Fortunately, reasoning about a single crashing testcase can be easier than
reasoning about the whole specification. This means that an LLM can often
usefully diagnose mistakes in \emph{hindsight} when presented with concrete
crashes, enabling automatic bug triage and specification repair. 

To close the loop, we deploy an LLM agent to automatically configure projects
for fuzzing in reproducible Docker containers. Together, these components
constitute \tool, an end-to-end bug finding system for C/C++ libraries,
operating on only a repository URL as input.

\newcommand{\agent}[1]{\raisebox{-0.05em}{\includegraphics[height=1em]{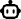}}~\textbf{#1}}
\newcommand{\prompt}[1]{\raisebox{-0.2em}{\includegraphics[height=1em]{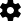}}~\textbf{#1}}

\subsection{Overview}
\label{sec:stitching-overview}

\begin{figure*}
  \centering
  \resizebox{\linewidth}{!}{
    \includegraphics{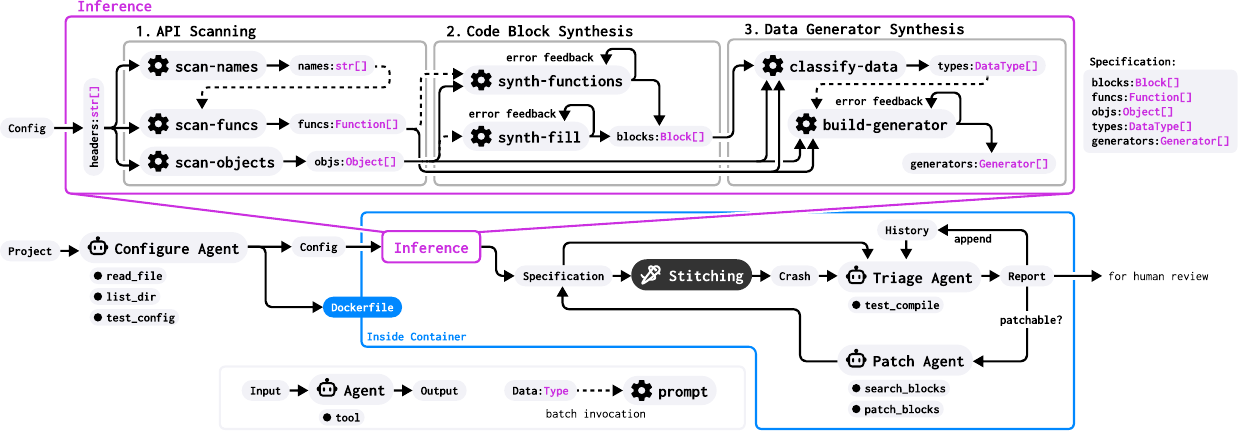}
  }
  \caption{The \tool workflow, from project configuration to bug finding.}
  \label{fig:llm}
\end{figure*}

\autoref{fig:llm} illustrates the \tool workflow. Here, an \agent{Agent} is a
component that can iterate and invoke tools before producing output. In
contrast, a \prompt{prompt} is invoked once (or in batches) and directly
produces output. 

Given a repository URL, \tool first configures the project for fuzzing
(\autoref{fig:llm}, bottom left) by
compiling the target with fuzz instrumentation and locating public header files (\agent{Configure Agent}).

Next, \tool synthesizes an initial specification (\autoref{fig:llm}, top). 
The \textbf{Inference Workflow} uses a multi-step prompt pipeline to synthesize a fuzzable specification from the project's public header files (\S\ref{sec:specification-synthesis}). It first scans the headers for API-relevant information, distilling constraints scattered throughout the code into a descriptive set of objects and functions that capture the library's behavior (\S\ref{sec:api-scanning}). For each library function, it synthesizes a code block, describing how to invoke the function, including the inputs and outputs it requires and produces. These code blocks are validated to ensure they compile (retrying if necessary) (\S\ref{sec:code-block-synthesis}). Finally, it analyzes the code blocks to identify custom data types that are used by the library and synthesizes generator functions for each to bootstrap the fuzzing process (\S\ref{sec:data-generator-synthesis}).

\looseness-1
\tool then enters an iterative fuzzing loop (\autoref{fig:llm}, bottom)
surrounding the core stitching implementation. Crashes are automatically triaged
by the \agent{Triage Agent} (\S\ref{sec:crash-triage}), which minimizes,
analyzes, and categorizes each crash. This includes identifying shallow bugs
that impede deeper exploration and likely false positives caused by incorrect
specifications. These cases trigger \emph{specification repair}
(\S\ref{sec:spec-repair}), where the \agent{Patch Agent} updates the
specification to preclude the undesirable crash. 
Fuzzing restarts with the updated specification.

This loop continues until a time budget is exhausted. The final output is a prioritized list of bug reports, each including a description, minimized testcase, crash report, and fully reproducible dockerized build.

\subsection{Project Configuration}
\label{sec:project-configuration}

For both reproducibility and sandboxing, \tool performs fuzzing inside isolated containers.
To configure a new target project, the \agent{Configure Agent} takes a
repository URL and produces two artifacts: a \texttt{Dockerfile} that builds the
target with fuzzer instrumentation and installs it to a fixed prefix, and a
\texttt{Config} JSON file listing the locations of public header files along
with any required linker flags or preprocessor definitions. As a validation
step, we require the agent to synthesize a small \texttt{test.cpp} program that
includes the identified headers and invokes a function from the target API. This
serves as a concrete check that the build succeeded, with usable headers. 

The agent is provided with a detailed prompt, the project's \texttt{README.md} (if one exists), and three tools: \texttt{read\_file} and \texttt{list\_dir} for navigating the repository, and \texttt{test\_configuration} for compiling a prospective \texttt{Dockerfile} and inspecting any errors.
Beyond the \texttt{test.cpp} validation, the process is best-effort: the agent may not discover every public header, particularly when they are spread across multiple directories or generated at build.
In practice, this means some functions may be absent from the specification. However, since missing headers simply reduce the scope of the specification rather than compromise its correctness, this is a graceful degradation.
When targeting a single project, a user can easily inspect and correct the header list in the \texttt{Config} file.
At scale (\textbf{RQ2}, \S\ref{sec:rq2}), the agent was robust despite its simplicity, successfully configuring roughly two-thirds of the 1365 targets without human involvement.
Subsequent pipeline steps run in the resulting container.

\subsection{Specification Synthesis}
\label{sec:specification-synthesis}

To synthesize a fuzzable specification from a project's public header files, \tool uses a multi-step prompt chaining pipeline (\autoref{fig:llm}, top half). Unlike the other tasks using an agentic approach, we found prompt chaining more effective here. Agents excel at open-ended exploration, but can produce inconsistent or partial results when exhaustively processing a fixed set of inputs. Prompt chaining provides more predictable, structured output for such tasks.

A core challenge in specification inference is that code blocks, while
synthesized individually, require shared context. Inputs and outputs must refer to the
same set of object types. A library may define \texttt{struct foo *} and also
typedef \texttt{foo\_t}; function signatures may use either (or even
\texttt{struct foo **}); yet code blocks must use a canonical
representation consistently. Similarly, extrinsic metadata must be referenced
consistently wherever an object is produced.

We therefore split specification synthesis into three phases. API Scanning (\S\ref{sec:api-scanning}) scans header files to distill the API into canonical descriptions of objects and functions. Code Block Synthesis (\S\ref{sec:code-block-synthesis}) uses these descriptions as ground truth to generate code blocks for each function. Finally, optionally, Data Generator Synthesis (\S\ref{sec:data-generator-synthesis})  identifies custom data types and generates bootstrapping functions for them.

\subsubsection{API Scanning}
\label{sec:api-scanning}

The first phase scans project headers to distill the API into canonical descriptions of functions (with \prompt{scan-funcs}) and objects (with \prompt{scan-objects}).

\noindent
\begin{minipage}[t]{0.48\columnwidth}
\begin{codebox}
Function:
  name: str
  signature: str
  description: str
  requirements: str[]
\end{codebox}
\vspace{0.5em}
\end{minipage}%
\hfill
\begin{minipage}[t]{0.48\columnwidth}
\begin{codebox}
Object:
  type: str
  aliases: str[]
  definition: str
  description: str
\end{codebox}
\vspace{0.5em}
\end{minipage}

This step achieves two key goals.
First, usage requirements for a function are often scattered across docstrings, object documentation, flag descriptions, and top-level comments about conventions.
API scanning distills this information into a compressed representation in the \texttt{requirements} field.
Second, this step produces a canonical set of objects to inform subsequent
synthesis steps. We explicitly prompt the LLM to group objects semantically, rather than syntactically, ensuring that e.g. \texttt{struct foo *}, \texttt{struct foo **} and \texttt{foo\_t} are all interpreted as the same object. One of these types is used as the canonical \texttt{type}, and the rest are recorded as \texttt{aliases}.

We treat code as plain text rather than using static analysis, reducing system complexity. In practice, the LLM is sufficiently effective at heuristically identifying public API functions to obviate precise syntax rules for the diversity of constructions, macros, and abstractions in real-world code. 
LLM hallucination risk is mitigated downstream by validating that code blocks compile and by repairing specifications that lead to spurious crashes (\S\ref{sec:spec-repair}). The system's key output---bug reports---are backed by dynamic evidence (minimized testcases) that attests to validity independent of LLM output.

We found that single prompts produced incomplete output for large codebases (100+ functions).  LLMs have finite output length limits and
attention quality
degrades over long sequences, causing later functions to receive less careful
treatment than earlier ones. Instead, we first synthesize a list of function names (\prompt{scan-names}), then batch them into groups of 20 for detailed synthesis (\prompt{scan-funcs}).

\subsubsection{Code Block Synthesis}
\label{sec:code-block-synthesis}

The second phase synthesizes a code block for each function description produced by API scanning. For each function, \prompt{synth-functions} generates a candidate code block, which is then validated to ensure it compiles and that its inputs and outputs consist only of canonical objects identified during scanning. If validation fails, the error is fed back to the model for up to three retry attempts. After all functions have been processed, some object types may lack constructors---either because the library expects callers to allocate them directly, or because synthesis failed for the relevant functions. We identify these \emph{unconstructable} types and use \prompt{synth-fill} to generate dedicated constructor blocks for them, following the same validate-and-retry loop.

Both prompts are provided with a detailed description of the fuzzing task and
the available FUZZ macros, as well as several examples of valid code blocks
(few-shot prompting is known to be effective for LLMs~\cite{fewshot}). We validate that generated code blocks have inputs and outputs belonging to the set of canonical objects, and that they compile successfully to catch any coding errors.
Validation errors are fed
back to the model up to three times before giving up on that code block. In
practice, this flow works quite well, with the LLM typically producing correct
code blocks on the first attempt, fewer than 1\% of code blocks fail validation
after three attempts.

\noindent
\begin{minipage}[t]{0.48\columnwidth}
\begin{codebox}
Block:
  name: str
  inputs: Arg[]
  outputs: Arg[]
  code: str
\end{codebox}
\vspace{0.5em}
\end{minipage}%
\hfill
\begin{minipage}[t]{0.48\columnwidth}
\begin{codebox}
Arg:
  name: str
  type: str
\end{codebox}
\vspace{0.5em}
\end{minipage}

\subsubsection{Data Generator Synthesis}
\label{sec:data-generator-synthesis}

Finally, the third phase builds data generator functions that seed the fuzzer with meaningful initial values for the \texttt{FUZZ\_PARAM} macros used in code blocks. Library APIs often consume data in specific, context-dependent formats: a string argument might be JSON data in \texttt{cJSON}, an attribute name in \texttt{pupnp}, or a comma-separated list of IP addresses in \texttt{c-ares}. Even within a single library, different functions can interpret the same data type in entirely different ways. While the fuzzing engine can gradually discover valid values through mutation alone, seeding it with well-formed examples significantly accelerates exploration.

We first use \prompt{classify-data} to scan the generated code blocks (along with their function descriptions) and produce natural language descriptions of the data formats each block consumes. Then, for each identified format, \prompt{build-generator} synthesizes a Python function that produces random values of the appropriate type. As before, we validate that each generator runs successfully, retrying with error feedback if needed. Errors in this phase are less consequential than in earlier phases: generators merely provide starting points, and the fuzzing engine remains free to mutate their output.

\noindent
\begin{minipage}[t]{0.48\columnwidth}
\begin{codebox}
DataType:
  name: str
  description: str
  used_by: str[]
\end{codebox}
\vspace{0.5em}
\end{minipage}%
\hfill
\begin{minipage}[t]{0.48\columnwidth}
\begin{codebox}
Generator:
  name: str
  code: str
\end{codebox}
\vspace{0.5em}
\end{minipage}

\subsection{Crash Triage}
\label{sec:crash-triage}

The stitching engine constructs and executes testcases to identify crashes. However, many crashes are duplicates or false positives resulting from specification misuse. While the stitching engine minimizes crashes by removing redundant block instances and exports them as standalone reproducers (\S\ref{sec:stitching-c-cpp}), these reproducers tend to be bloated.

To better triage crashes, \tool uses the \agent{Triage Agent} to progressively \emph{distill} crash reproducers down to minimal, self-contained testcases while analyzing the bug. Triage also flags crashes that appear to be spurious, informing specification refinement (\S\ref{sec:spec-repair}).
The \agent{Triage Agent} receives reproducers along with the relevant specification context. Its primary tool is \texttt{test\_compile}, which allows the agent to compile and run modified testcases and inspect sanitizer reports. The agent is instructed to systematically reduce the reproducer by removing unnecessary code, simplifying arguments, and performing A/B tests to isolate which aspects of the testcase are necessary to trigger the bug.

After investigation, the agent produces a report detailing the predicted root cause, the minimized reproducer, a human-readable description, and categorical features such as bug type, degree of attacker control, and trigger surface. Based on these attributes, the agent assigns a severity of \textbf{p1} (security), \textbf{p2} (stability), or \textbf{p3} (misuse/false positive), and determines whether the specification can be patched to avoid the crash, informing the \agent{Patch Agent} (\S\ref{sec:spec-repair}).

Finally, the agent maintains a history of all generated reports to shortcut triage of similar crashes and perform semantic deduplication. Bugs that share the same root cause as a previously triaged crash are marked as duplicates.

\subsection{Specification Repair}
\label{sec:spec-repair}

Given a triaged crash, the \agent{Patch Agent} updates the specification, for
two reasons. First, true bugs may be shallow, crashing frequently and blocking the fuzzer from exploring for deeper bugs. Second, inferred specifications can be
incorrect or incomplete, leading to spurious crashes that do not reflect valid API usage.
In early experiments, we found that incorrect specifications often lead to an entire family of related bugs. Rather than patching each on a case-by-case basis, we instruct the agent to generalize from its report and patch the specification for the entire family.

The \agent{Patch Agent} is provided with two tools: \texttt{search\_blocks} (regex search over specification code blocks) and \texttt{patch\_blocks} (edit code within blocks). Newly-patched blocks are validated to ensure they compile. If successful, we migrate the fuzz corpus, dropping testcases that referenced patched blocks, and restart fuzzing with the new specification.

\section{Evaluation}
\label{sec:evaluation}

This section addresses the following research questions:
\begin{itemize}
  \item \textbf{RQ1 (SOTA Evaluation)}: How does \tool compare to other SOTA systems at code coverage and bug-finding ability?
  \item \textbf{RQ2 (Real-World Automatic Bug Finding)}: How effective is \tool at autonomous bug finding when deployed in the wild?
\end{itemize}
To answer them, we evaluated \tool both on a suite of well-tested benchmarks and on a large-scale real-world case study of 1365 open-source projects. \tool obtains significantly higher coverage and finds more bugs (with more precision) than other SOTA systems, while being faster and more cost-effective (in terms of \$/bug). \tool is also capable of real-world fully autonomous bug finding; to date, we have reported 131 new bugs across 102 projects, \numpatched of which have already been patched.

\subsection{Experimental Setup}

\customsection{Benchmarks}
For the head-to-head experiment, we evaluate \tool on a superset of 33 benchmarks used by prior work, consisting of both C and C++ targets across a variety of domains. For extremely large benchmarks, we limit the scope to a few select headers to allow other tools to run with a reasonable budget. We use project commits between 1-4 years old (see \autoref{table:cov-results}) to avoid favoring \tool over competitors who previously discovered bugs that may have since been fixed.

\customsection{Tools}
We evaluate against three recent SOTA tools: \work{PromeFuzz}~\cite{promefuzz}, \work{PromptFuzz}~\cite{promptfuzz}, and \work{OGHarn}~\cite{ogharn}, as well as manually-written fuzz harnesses from the benchmarks or \work{OSS-Fuzz} (collectively referred to as \work{OSS-Fuzz}). Each tool uses the same fuzzer runtime (\work{LibAFL} \texttt{0.14.1} in forkserver mode) and LLM model (\texttt{gpt-5-2025-08-07})~\footnote{except for \work{PromeFuzz} where this was cost-prohibitive. Instead, we use \texttt{DeepSeek-V3.2-Exp} in conjunction with \texttt{mxbai-embed-large} for embeddings at the author's recommendation}.

\customsection{Runtime}
Each tool is allocated 1vCPU and 2GB RAM. The coverage experiments are performed on a Google Cloud Kubernetes cluster using c2-standard-8 nodes with Intel Cascade Lake processors. We ran each tool to generate the harness or specification once, then ran the resulting artifact 5 times independently for 24 hours each. For the head-to-head experiment, we ran \tool without the \agent{Patch Agent}, using only the initially-generated specification to measure just the coverage of the initial specification and force equal resource consumption (other tools similarly do not use LLMs at runtime).

\customsection{Coverage Measurements}
We measure code coverage by rerunning each tool on the resulting corpus, using an identical coverage-instrumented build of each benchmark, excluding coverage of tool-specific harness code.

\customsection{Bug Discovery}
After fuzzing, we aggregated all crashes and automatically triaged them where supported (\tool's \agent{Triage Agent} and \work{PromeFuzz}'s crash analyzer), filtering out detected false positives (labeled \textbf{X} in the tables). We
then manually deduplicated remaining bugs by crash location and root cause, and
categorized each as a true positive (\textbf{TP})---a genuine bug in the
library---or a false positive (\textbf{FP})---a crash caused by invalid API
usage in the harness or specification.

\subsection{RQ1: SOTA Evaluation}
\label{sec:rq1}

\newcommand{\na}{\crossbox[black!20]}

\definecolor{bestgreen}{HTML}{E1BEE7}
\newcommand{\best}[1]{\textbf{#1}}
\newcommand{\bestsig}[1]{\cellcolor{bestgreen}\textbf{#1}}

\definecolor{heatyellow}{HTML}{FFE082}
\definecolor{heatgreen}{HTML}{A5D6A7}
\definecolor{heatred}{HTML}{EF9A9A}

\autoref{table:cov-results} shows coverage and bug discovery results.

\looseness-1
\customsection{Coverage}
Section \textbf{(A)} shows arithmetic mean line coverage per tool, with the highest per target in \textbf{bold}.
Following best practices~\cite{schloegel2024sok}, we use the Mann-Whitney U Test for statistical significance and \colorbox{bestgreen}{\textbf{shade}} the best value if it beats the other results with $p < 0.05$.
Overall, \tool achieves the strongest coverage, with statistically significant wins on 21/33 benchmarks. It beats \work{OGHarn} on every benchmark, \work{OSS-Fuzz} on 18/22, \work{PromptFuzz} on 10/12, and \work{PromeFuzz} on 16/22.

\customsection{Bug Finding}
Section \textbf{(B)} shows unique bugs found by each tool. Columns labeled
\textbf{X} indicate crashes that were automatically filtered out by a tool's
triage system before manual review: for \tool, these are crashes assigned
priority \textbf{p3} (predicted misuse); for \work{PromeFuzz}, those labeled
\textbf{F} by its crash analyzer. The remaining crashes were then manually
deduplicated and categorized as true positives (\textbf{TP}) or false positives
(\textbf{FP}).
\tool found 30 unique bugs on this benchmark suite, compared to 10 found by all other tools combined. Of these, 15 were \emph{previously unknown} bugs still present in the latest versions, which we reported to maintainers; 9 have already been patched.

While the total bug count is dominated by the buggy projects \texttt{libucl} and \texttt{libfyaml}, \tool found bugs in 14 unique projects. In \texttt{libaom}, despite having the lowest coverage, \tool was the only tool to find a true positive bug. \work{PromeFuzz} and \work{PromptFuzz} had many false positives in \texttt{libaom} due to misuse of the variadic \texttt{aom\_codec\_ctrl} function, which also inflated their coverage by triggering handler functions without proper argument constraints. 

\customsection{Bug Finding Precision}
\work{OGHarn} had the highest precision (100\%), but this reflects its conservative design. It restricts fuzzing to functions accepting bytestring arguments and discards harnesses that crash early, so reported crashes tend to be data-dependent true positives. \work{OSS-Fuzz} harnesses found no bugs on these benchmarks.
Among LLM-based tools, \tool had the highest precision (70\%, 30 true positives), ahead of \work{PromeFuzz} (12\%, 5 TPs) and \work{PromptFuzz} (2\%, 1 TP). \tool's false positives were less severe (marked \textbf{p2} rather than \textbf{p1}) and typically occurred in less well-documented APIs.

The precision gap is partly explained by triage differences. \work{PromeFuzz}'s analyzer was conservative, discarding only 8 of 50 crashes and overlooking common misuse patterns like null pointer passing. \tool's \agent{Triage Agent} was more aggressive, iteratively compiling and running modified testcases to test hypotheses, filtering 221 of 265 crashes. We observed no obvious false negatives in either system.

\newcommand{\heatcellb}[4]{%
  \pgfmathsetmacro{\pct}{round((#1 - #2) / (#3 - #2) * 100)}%
  \edef\temp{\noexpand\cellcolor{#4!\pct}}%
  \temp%
}

\ExplSyntaxOn
\cs_new_protected:Nn \__pantigny_crossbox:nnn
 {
 \tikz \draw [ #3 ]
 ( #1 -| \inteval { #2 + 1 } ) -- ( \inteval { #1 + 1 } -| #2 )
 ( #1 -| #2 ) -- ( \inteval { #1 + 1 } -| \inteval { #2 + 1 } ) ;
 }
\NewDocumentCommand \crossbox { ! O { } }
 {
 \tl_gput_right:Ne \g_nicematrix_code_before_tl
 {
 \__pantigny_crossbox:nnn
 { \int_use:c { c@iRow } }
 { \int_use:c { c@jCol } }
 { \exp_not:n { #1 } }
 }
 \ignorespaces
 }
\ExplSyntaxOff

\begin{table*}
  \centering
  \footnotesize
  \begin{NiceTabular}[color-inside]{|[start=4]Xcr|[start=4]@{}w{c}{4pt}@{}|[start=4]r|[start=4]r|[start=4]r|[start=4]r|[start=4]r|[start=4]@{}w{c}{4pt}@{}|[start=3]w{c}{0.7em}w{c}{0.7em}w{c}{0.7em}|[start=3]w{c}{0.7em}w{c}{0.7em}w{c}{0.7em}|[start=3]w{c}{0.7em}w{c}{0.7em}|[start=3]w{c}{0.7em}w{c}{0.7em}|[start=3]w{c}{0.7em}w{c}{0.7em}|[start=3]w{c}{0.7em}w{c}{0.7em}|[start=3]}
    & & & & \Block{1-5}{\textbf{(A) Line Coverage \%}} & & & & & & \Block{1-12}{\textbf{(B) Unique Bugs Found}} \\

    & & & & \Block[c]{2-1}{\circled{1}} & \Block[c]{2-1}{\circled{2}} & \Block[c]{2-1}{\circled{3}} & \Block[c]{2-1}{\circled{4}} & \Block[c]{2-1}{\circled{5}}
    & & \Block[c]{1-3}{\circled{1}} & & & \Block[c]{1-3}{\circled{2}} & & & \Block[c]{1-2}{\circled{3}} & &\Block[c]{1-2}{\circled{4}} & & \Block[c]{1-2}{\circled{5}} \\

    \cline{11-22}
  
    \textbf{Benchmark} & \textbf{Commit} & \textbf{API} & & & & & & & & \textbf{X} & \textbf{FP} & \textbf{TP} & \textbf{X} & \textbf{FP} & \textbf{TP} & \textbf{FP} & \textbf{TP} & \textbf{FP} & \textbf{TP} & \textbf{FP} & \textbf{TP} \\
    \cline{1-3} \cline{5-9} \cline{11-22}
    \texttt{c-ares} & \texttt{a8c0917} (1.2) & 136 \CIRCLE& & 57.8 & \best{57.9} & 55.8 & 29.6 & 34.4&& \heatcellb{1}{0}{56}{heatyellow}1 & \heatcellb{0}{0}{24}{heatred}0 & \heatcellb{0}{0}{10}{heatgreen}0& \heatcellb{0}{0}{56}{heatyellow}0 & \heatcellb{1}{0}{24}{heatred}1 & \heatcellb{0}{0}{10}{heatgreen}0& \heatcellb{0}{0}{24}{heatred}0 & \heatcellb{0}{0}{10}{heatgreen}0& \heatcellb{0}{0}{24}{heatred}0 & \heatcellb{0}{0}{10}{heatgreen}0& \heatcellb{0}{0}{24}{heatred}0 & \heatcellb{0}{0}{10}{heatgreen}0 \\
\texttt{cjson} & \texttt{12c4bf1} (1.4) & 78 \CIRCLE& & \bestsig{85.2} & 82.7 & 82.2 & 58.2 & 40.6&& \heatcellb{3}{0}{56}{heatyellow}3 & \heatcellb{0}{0}{24}{heatred}0 & \heatcellb{0}{0}{10}{heatgreen}0& \heatcellb{0}{0}{56}{heatyellow}0 & \heatcellb{0}{0}{24}{heatred}0 & \heatcellb{0}{0}{10}{heatgreen}0& \heatcellb{2}{0}{24}{heatred}2 & \heatcellb{0}{0}{10}{heatgreen}0& \heatcellb{0}{0}{24}{heatred}0 & \heatcellb{0}{0}{10}{heatgreen}0& \heatcellb{0}{0}{24}{heatred}0 & \heatcellb{0}{0}{10}{heatgreen}0 \\
\texttt{lcms} & \texttt{676e803} (1.2) & 358 \CIRCLE& & \bestsig{54.3} & 46.3 & 37.7 & 11.5 & 18.0&& \heatcellb{6}{0}{56}{heatyellow}6 & \heatcellb{1}{0}{24}{heatred}1 & \heatcellb{0}{0}{10}{heatgreen}0& \heatcellb{3}{0}{56}{heatyellow}3 & \heatcellb{6}{0}{24}{heatred}6 & \heatcellb{0}{0}{10}{heatgreen}0& \heatcellb{1}{0}{24}{heatred}1 & \heatcellb{0}{0}{10}{heatgreen}0& \heatcellb{0}{0}{24}{heatred}0 & \heatcellb{0}{0}{10}{heatgreen}0& \heatcellb{0}{0}{24}{heatred}0 & \heatcellb{0}{0}{10}{heatgreen}0 \\
\texttt{libmagic} & \texttt{0fa2c8c} (1.3) & 18 \CIRCLE& & \best{52.9} & 49.9 & 49.4 & 39.1 & \na && \heatcellb{0}{0}{56}{heatyellow}0 & \heatcellb{0}{0}{24}{heatred}0 & \heatcellb{1}{0}{10}{heatgreen}1& \heatcellb{0}{0}{56}{heatyellow}0 & \heatcellb{0}{0}{24}{heatred}0 & \heatcellb{0}{0}{10}{heatgreen}0& \heatcellb{0}{0}{24}{heatred}0 & \heatcellb{1}{0}{10}{heatgreen}1& \heatcellb{0}{0}{24}{heatred}0 & \heatcellb{0}{0}{10}{heatgreen}0& \heatcellb{0}{0}{24}{heatred}0 & \heatcellb{0}{0}{10}{heatgreen}0 \\
\texttt{libpcap} & \texttt{e17fe06} (1.2) & 188 \CIRCLE& & \bestsig{47.4} & 35.5 & 29.0 & 22.7 & 24.6&& \heatcellb{6}{0}{56}{heatyellow}6 & \heatcellb{0}{0}{24}{heatred}0 & \heatcellb{1}{0}{10}{heatgreen}1& \heatcellb{0}{0}{56}{heatyellow}0 & \heatcellb{0}{0}{24}{heatred}0 & \heatcellb{0}{0}{10}{heatgreen}0& \heatcellb{0}{0}{24}{heatred}0 & \heatcellb{0}{0}{10}{heatgreen}0& \heatcellb{0}{0}{24}{heatred}0 & \heatcellb{0}{0}{10}{heatgreen}0& \heatcellb{0}{0}{24}{heatred}0 & \heatcellb{0}{0}{10}{heatgreen}0 \\
\texttt{sqlite3} & \texttt{0df847c} (1.2) & 317 \CIRCLE& & \bestsig{36.2} & 29.5 & 13.4 & 13.2 & 6.2&& \heatcellb{3}{0}{56}{heatyellow}3 & \heatcellb{0}{0}{24}{heatred}0 & \heatcellb{1}{0}{10}{heatgreen}1& \heatcellb{2}{0}{56}{heatyellow}2 & \heatcellb{4}{0}{24}{heatred}4 & \heatcellb{0}{0}{10}{heatgreen}0& \heatcellb{3}{0}{24}{heatred}3 & \heatcellb{0}{0}{10}{heatgreen}0& \heatcellb{0}{0}{24}{heatred}0 & \heatcellb{0}{0}{10}{heatgreen}0& \heatcellb{0}{0}{24}{heatred}0 & \heatcellb{0}{0}{10}{heatgreen}0 \\
\texttt{zlib} & \texttt{ef24c4c} (1.2) & 82 \CIRCLE& & \best{85.8} & 84.1 & 83.0 & 45.7 & 72.1&& \heatcellb{4}{0}{56}{heatyellow}4 & \heatcellb{0}{0}{24}{heatred}0 & \heatcellb{0}{0}{10}{heatgreen}0& \heatcellb{1}{0}{56}{heatyellow}1 & \heatcellb{1}{0}{24}{heatred}1 & \heatcellb{0}{0}{10}{heatgreen}0& \heatcellb{4}{0}{24}{heatred}4 & \heatcellb{0}{0}{10}{heatgreen}0& \heatcellb{0}{0}{24}{heatred}0 & \heatcellb{0}{0}{10}{heatgreen}0& \heatcellb{1}{0}{24}{heatred}1 & \heatcellb{0}{0}{10}{heatgreen}0 \\
\cline{1-3} \cline{5-9} \cline{11-22}
\texttt{libaom} & \texttt{99fcd81} (1.0) & 56 \CIRCLE& & 15.4 & 41.1 & 33.7 & \na  & \bestsig{58.2}&& \heatcellb{2}{0}{56}{heatyellow}2 & \heatcellb{1}{0}{24}{heatred}1 & \heatcellb{1}{0}{10}{heatgreen}1& \heatcellb{0}{0}{56}{heatyellow}0 & \heatcellb{6}{0}{24}{heatred}6 & \heatcellb{0}{0}{10}{heatgreen}0& \heatcellb{24}{0}{24}{heatred}24 & \heatcellb{0}{0}{10}{heatgreen}0& \na & \na& \heatcellb{0}{0}{24}{heatred}0 & \heatcellb{0}{0}{10}{heatgreen}0 \\
\texttt{libjpeg-turbo} & \texttt{36ac5b8} (1.0) & 148 \CIRCLE& & \bestsig{61.9} & 30.8 & 28.8 & \na  & 31.5&& \heatcellb{4}{0}{56}{heatyellow}4 & \heatcellb{1}{0}{24}{heatred}1 & \heatcellb{1}{0}{10}{heatgreen}1& \heatcellb{0}{0}{56}{heatyellow}0 & \heatcellb{3}{0}{24}{heatred}3 & \heatcellb{0}{0}{10}{heatgreen}0& \heatcellb{5}{0}{24}{heatred}5 & \heatcellb{0}{0}{10}{heatgreen}0& \na & \na& \heatcellb{0}{0}{24}{heatred}0 & \heatcellb{0}{0}{10}{heatgreen}0 \\
\texttt{libpng} & \texttt{738f5e7} (1.0) & 257 \CIRCLE& & \bestsig{52.1} & 30.0 & 4.7 & \na  & 0.0&& \heatcellb{9}{0}{56}{heatyellow}9 & \heatcellb{0}{0}{24}{heatred}0 & \heatcellb{0}{0}{10}{heatgreen}0& \heatcellb{0}{0}{56}{heatyellow}0 & \heatcellb{1}{0}{24}{heatred}1 & \heatcellb{0}{0}{10}{heatgreen}0& \heatcellb{1}{0}{24}{heatred}1 & \heatcellb{0}{0}{10}{heatgreen}0& \na & \na& \heatcellb{0}{0}{24}{heatred}0 & \heatcellb{0}{0}{10}{heatgreen}0 \\
\texttt{libtiff} & \texttt{fcd4c86} (1.0) & 200 \CIRCLE& & \best{32.9} & 31.3 & 12.5 & \na  & 24.6&& \heatcellb{34}{0}{56}{heatyellow}34 & \heatcellb{1}{0}{24}{heatred}1 & \heatcellb{1}{0}{10}{heatgreen}1& \heatcellb{1}{0}{56}{heatyellow}1 & \heatcellb{0}{0}{24}{heatred}0 & \heatcellb{0}{0}{10}{heatgreen}0& \heatcellb{0}{0}{24}{heatred}0 & \heatcellb{0}{0}{10}{heatgreen}0& \na & \na& \heatcellb{0}{0}{24}{heatred}0 & \heatcellb{0}{0}{10}{heatgreen}0 \\
\texttt{libvpx} & \texttt{9514ab6} (1.0) & 48 \CIRCLE& & 5.4 & \bestsig{38.5} & 13.7 & \na  & 9.8&& \heatcellb{4}{0}{56}{heatyellow}4 & \heatcellb{0}{0}{24}{heatred}0 & \heatcellb{0}{0}{10}{heatgreen}0& \heatcellb{0}{0}{56}{heatyellow}0 & \heatcellb{10}{0}{24}{heatred}10 & \heatcellb{0}{0}{10}{heatgreen}0& \heatcellb{1}{0}{24}{heatred}1 & \heatcellb{0}{0}{10}{heatgreen}0& \na & \na& \heatcellb{0}{0}{24}{heatred}0 & \heatcellb{0}{0}{10}{heatgreen}0 \\
\cline{1-3} \cline{5-9} \cline{11-22}
\texttt{curl} & \texttt{4c50998} (1.0) & 100 \CIRCLE& & 17.6 & \bestsig{27.4} & \na  & \na  & 3.1&& \heatcellb{2}{0}{56}{heatyellow}2 & \heatcellb{0}{0}{24}{heatred}0 & \heatcellb{0}{0}{10}{heatgreen}0& \heatcellb{0}{0}{56}{heatyellow}0 & \heatcellb{0}{0}{24}{heatred}0 & \heatcellb{0}{0}{10}{heatgreen}0& \na & \na& \na & \na& \heatcellb{0}{0}{24}{heatred}0 & \heatcellb{0}{0}{10}{heatgreen}0 \\
\texttt{liblouis} & \texttt{3d95765} (1.1) & 30 \CIRCLE& & \bestsig{72.5} & 51.0 & \na  & \na  & 0.0&& \heatcellb{4}{0}{56}{heatyellow}4 & \heatcellb{2}{0}{24}{heatred}2 & \heatcellb{1}{0}{10}{heatgreen}1& \heatcellb{0}{0}{56}{heatyellow}0 & \heatcellb{0}{0}{24}{heatred}0 & \heatcellb{2}{0}{10}{heatgreen}2& \na & \na& \na & \na& \heatcellb{0}{0}{24}{heatred}0 & \heatcellb{0}{0}{10}{heatgreen}0 \\
\texttt{exiv2} \textcolor{orange}{\footnotesize \texttt{C++}} & \texttt{04e1ea3} (1.0) & 939 \CIRCLE& & \bestsig{35.1} & 30.0 & \na  & \na  & 9.0&& \heatcellb{11}{0}{56}{heatyellow}11 & \heatcellb{0}{0}{24}{heatred}0 & \heatcellb{0}{0}{10}{heatgreen}0& \heatcellb{0}{0}{56}{heatyellow}0 & \heatcellb{4}{0}{24}{heatred}4 & \heatcellb{0}{0}{10}{heatgreen}0& \na & \na& \na & \na& \heatcellb{0}{0}{24}{heatred}0 & \heatcellb{0}{0}{10}{heatgreen}0 \\
\texttt{pugixml} \textcolor{orange}{\footnotesize \texttt{C++}} & \texttt{06318b0} (1.1) & 342 \CIRCLE& & \best{74.7} & 73.2 & \na  & \na  & 42.4&& \heatcellb{6}{0}{56}{heatyellow}6 & \heatcellb{0}{0}{24}{heatred}0 & \heatcellb{0}{0}{10}{heatgreen}0& \heatcellb{0}{0}{56}{heatyellow}0 & \heatcellb{0}{0}{24}{heatred}0 & \heatcellb{0}{0}{10}{heatgreen}0& \na & \na& \na & \na& \heatcellb{0}{0}{24}{heatred}0 & \heatcellb{0}{0}{10}{heatgreen}0 \\
\texttt{re2} & \texttt{c84a140} (0.9) & 97 \CIRCLE& & \bestsig{76.1} & 62.9 & \na  & \na  & \na && \heatcellb{1}{0}{56}{heatyellow}1 & \heatcellb{1}{0}{24}{heatred}1 & \heatcellb{1}{0}{10}{heatgreen}1& \heatcellb{0}{0}{56}{heatyellow}0 & \heatcellb{0}{0}{24}{heatred}0 & \heatcellb{0}{0}{10}{heatgreen}0& \na & \na& \na & \na& \heatcellb{0}{0}{24}{heatred}0 & \heatcellb{0}{0}{10}{heatgreen}0 \\
\texttt{tinygltf} \textcolor{orange}{\footnotesize \texttt{C++}} & \texttt{a5e653e} (1.0) & 58 \CIRCLE& & \bestsig{46.9} & 27.9 & \na  & \na  & 11.1&& \heatcellb{1}{0}{56}{heatyellow}1 & \heatcellb{0}{0}{24}{heatred}0 & \heatcellb{0}{0}{10}{heatgreen}0& \heatcellb{0}{0}{56}{heatyellow}0 & \heatcellb{0}{0}{24}{heatred}0 & \heatcellb{0}{0}{10}{heatgreen}0& \na & \na& \na & \na& \heatcellb{0}{0}{24}{heatred}0 & \heatcellb{0}{0}{10}{heatgreen}0 \\
\cline{1-3} \cline{5-9} \cline{11-22}
\texttt{cgltf} & \texttt{de39988} (2.0) & 38 \CIRCLE& & \bestsig{54.7} & \na  & \na  & 0.4 & 7.2&& \heatcellb{6}{0}{56}{heatyellow}6 & \heatcellb{0}{0}{24}{heatred}0 & \heatcellb{0}{0}{10}{heatgreen}0& \na & \na & \na& \na & \na& \heatcellb{0}{0}{24}{heatred}0 & \heatcellb{0}{0}{10}{heatgreen}0& \heatcellb{0}{0}{24}{heatred}0 & \heatcellb{0}{0}{10}{heatgreen}0 \\
\texttt{hdf5} & \texttt{0394b03} (1.8) & 102 \LEFTcircle& & \bestsig{16.5} & \na  & \na  & 5.6 & 8.5&& \heatcellb{13}{0}{56}{heatyellow}13 & \heatcellb{0}{0}{24}{heatred}0 & \heatcellb{0}{0}{10}{heatgreen}0& \na & \na & \na& \na & \na& \heatcellb{0}{0}{24}{heatred}0 & \heatcellb{0}{0}{10}{heatgreen}0& \heatcellb{0}{0}{24}{heatred}0 & \heatcellb{0}{0}{10}{heatgreen}0 \\
\texttt{libical} & \texttt{460b8d7} (1.8) & 280 \LEFTcircle& & \bestsig{22.0} & \na  & \na  & 1.9 & 2.1&& \heatcellb{2}{0}{56}{heatyellow}2 & \heatcellb{0}{0}{24}{heatred}0 & \heatcellb{1}{0}{10}{heatgreen}1& \na & \na & \na& \na & \na& \heatcellb{0}{0}{24}{heatred}0 & \heatcellb{0}{0}{10}{heatgreen}0& \heatcellb{0}{0}{24}{heatred}0 & \heatcellb{0}{0}{10}{heatgreen}0 \\
\texttt{libucl} & \texttt{51c5e2f} (1.9) & 141 \CIRCLE& & \bestsig{64.3} & \na  & \na  & 37.8 & 25.7&& \heatcellb{16}{0}{56}{heatyellow}16 & \heatcellb{0}{0}{24}{heatred}0 & \heatcellb{10}{0}{10}{heatgreen}10& \na & \na & \na& \na & \na& \heatcellb{0}{0}{24}{heatred}0 & \heatcellb{2}{0}{10}{heatgreen}2& \heatcellb{0}{0}{24}{heatred}0 & \heatcellb{0}{0}{10}{heatgreen}0 \\
\texttt{openexr} \textcolor{orange}{\footnotesize \texttt{C++}} & \texttt{d669510} (1.8) & 87 \LEFTcircle& & 1.5 & \na  & \na  & 0.0 & \bestsig{2.9}&& \heatcellb{5}{0}{56}{heatyellow}5 & \heatcellb{0}{0}{24}{heatred}0 & \heatcellb{0}{0}{10}{heatgreen}0& \na & \na & \na& \na & \na& \heatcellb{0}{0}{24}{heatred}0 & \heatcellb{0}{0}{10}{heatgreen}0& \heatcellb{0}{0}{24}{heatred}0 & \heatcellb{0}{0}{10}{heatgreen}0 \\
\texttt{pcre2} & \texttt{a678783} (1.2) & 78 \CIRCLE& & 52.5 & \na  & \na  & 32.9 & \bestsig{71.4}&& \heatcellb{6}{0}{56}{heatyellow}6 & \heatcellb{0}{0}{24}{heatred}0 & \heatcellb{0}{0}{10}{heatgreen}0& \na & \na & \na& \na & \na& \heatcellb{0}{0}{24}{heatred}0 & \heatcellb{0}{0}{10}{heatgreen}0& \heatcellb{0}{0}{24}{heatred}0 & \heatcellb{0}{0}{10}{heatgreen}0 \\
\cline{1-3} \cline{5-9} \cline{11-22}
\texttt{faup} & \texttt{3a26d0a} (2.8) & 45 \LEFTcircle& & \bestsig{10.8} & \na  & \na  & 9.8 & \na && \heatcellb{1}{0}{56}{heatyellow}1 & \heatcellb{0}{0}{24}{heatred}0 & \heatcellb{0}{0}{10}{heatgreen}0& \na & \na & \na& \na & \na& \heatcellb{0}{0}{24}{heatred}0 & \heatcellb{1}{0}{10}{heatgreen}1& \na & \na \\
\texttt{geos} & \texttt{c8b889b} (1.8) & 140 \CIRCLE& & \bestsig{45.7} & \na  & \na  & 0.6 & \na && \heatcellb{3}{0}{56}{heatyellow}3 & \heatcellb{1}{0}{24}{heatred}1 & \heatcellb{1}{0}{10}{heatgreen}1& \na & \na & \na& \na & \na& \heatcellb{0}{0}{24}{heatred}0 & \heatcellb{0}{0}{10}{heatgreen}0& \na & \na \\
\texttt{lexbor} & \texttt{6c219fe} (1.8) & 47 \LEFTcircle& & \bestsig{33.5} & \na  & \na  & 13.1 & \na && \heatcellb{5}{0}{56}{heatyellow}5 & \heatcellb{1}{0}{24}{heatred}1 & \heatcellb{0}{0}{10}{heatgreen}0& \na & \na & \na& \na & \na& \heatcellb{0}{0}{24}{heatred}0 & \heatcellb{0}{0}{10}{heatgreen}0& \na & \na \\
\texttt{libfyaml} & \texttt{1f520e6} (2.3) & 170 \CIRCLE& & \bestsig{58.6} & \na  & \na  & 7.1 & \na && \heatcellb{56}{0}{56}{heatyellow}56 & \heatcellb{4}{0}{24}{heatred}4 & \heatcellb{5}{0}{10}{heatgreen}5& \na & \na & \na& \na & \na& \heatcellb{0}{0}{24}{heatred}0 & \heatcellb{0}{0}{10}{heatgreen}0& \na & \na \\
\texttt{stormlib} & \texttt{6052223} (1.8) & 88 \CIRCLE& & \bestsig{49.6} & \na  & \na  & 2.8 & \na && \heatcellb{6}{0}{56}{heatyellow}6 & \heatcellb{0}{0}{24}{heatred}0 & \heatcellb{3}{0}{10}{heatgreen}3& \na & \na & \na& \na & \na& \heatcellb{0}{0}{24}{heatred}0 & \heatcellb{1}{0}{10}{heatgreen}1& \na & \na \\
\cline{1-3} \cline{5-9} \cline{11-22}
\texttt{ngiflib} & \texttt{db19270} (2.1) & 7 \CIRCLE& & \bestsig{86.3} & 19.2 & \na  & \na  & \na && \heatcellb{0}{0}{56}{heatyellow}0 & \heatcellb{0}{0}{24}{heatred}0 & \heatcellb{0}{0}{10}{heatgreen}0& \heatcellb{0}{0}{56}{heatyellow}0 & \heatcellb{0}{0}{24}{heatred}0 & \heatcellb{2}{0}{10}{heatgreen}2& \na & \na& \na & \na& \na & \na \\
\texttt{ffjpeg} & \texttt{caade60} (4.1) & 40 \CIRCLE& & 92.9 & \bestsig{95.3} & \na  & \na  & \na && \heatcellb{0}{0}{56}{heatyellow}0 & \heatcellb{0}{0}{24}{heatred}0 & \heatcellb{0}{0}{10}{heatgreen}0& \heatcellb{1}{0}{56}{heatyellow}1 & \heatcellb{0}{0}{24}{heatred}0 & \heatcellb{1}{0}{10}{heatgreen}1& \na & \na& \na & \na& \na & \na \\
\texttt{loguru} \textcolor{orange}{\footnotesize \texttt{C++}} & \texttt{4adaa18} (2.8) & 87 \CIRCLE& & 65.8 & \bestsig{78.9} & \na  & \na  & \na && \heatcellb{1}{0}{56}{heatyellow}1 & \heatcellb{0}{0}{24}{heatred}0 & \heatcellb{0}{0}{10}{heatgreen}0& \heatcellb{0}{0}{56}{heatyellow}0 & \heatcellb{1}{0}{24}{heatred}1 & \heatcellb{0}{0}{10}{heatgreen}0& \na & \na& \na & \na& \na & \na \\
\texttt{rapidcsv} \textcolor{orange}{\footnotesize \texttt{C++}} & \texttt{083851d} (1.0) & 27 \CIRCLE& & \bestsig{90.8} & 80.2 & \na  & \na  & \na && \heatcellb{0}{0}{56}{heatyellow}0 & \heatcellb{0}{0}{24}{heatred}0 & \heatcellb{2}{0}{10}{heatgreen}2& \heatcellb{0}{0}{56}{heatyellow}0 & \heatcellb{0}{0}{24}{heatred}0 & \heatcellb{0}{0}{10}{heatgreen}0& \na & \na& \na & \na& \na & \na \\
    \cline{1-3} \cline{5-9} \cline{11-22}
  \end{NiceTabular}
  \caption[Head-to-head benchmark results showing arithmetic mean line coverage and number of unique bugs found, computed over 5 runs for 24 hours. Commits show age in years relative to the time of our evaluation. API \CIRCLE: full scope, \LEFTcircle: partial scope. In the coverage subtable, \textbf{bold}: best value; \colorbox{bestgreen}{\textbf{shaded}} indicates the best value was better than each other result with $p < 0.05$ (Mann-Whitney U Test). In the bugs subtable, \textbf{X}: filtered out by bug analyzer, \textbf{FP}: false positive bug, \textbf{TP}: true positive bug, values are shaded according to value. (1): \tool, (2): \work{PromeFuzz}, (3): \work{PromptFuzz}, (4): \work{OGHarn}, (5): \work{OSS-Fuzz}]{Head-to-head benchmark results showing arithmetic mean line coverage and number of unique bugs found, computed over 5 runs for 24 hours. Commits show age in years relative to the time of our evaluation. API \CIRCLE: full scope, \LEFTcircle: partial scope. In the coverage subtable, \textbf{bold}: best value; \colorbox{bestgreen}{\textbf{shaded}} indicates the best value was better than each other result with $p < 0.05$ (Mann-Whitney U Test). In the bugs subtable, \textbf{X}: filtered out by bug analyzer, \textbf{FP}: false positive bug, \textbf{TP}: true positive bug, values are shaded according to value. \circled{1}: \tool, \circled{2}: \work{PromeFuzz}, \circled{3}: \work{PromptFuzz}, \circled{4}: \work{OGHarn}, \circled{5}: \work{OSS-Fuzz}}
  \label{table:cov-results}
\end{table*}

\looseness-1
\customsection{Speed and Cost}
\tool was the fastest tool to infer specifications, averaging 19 minutes per benchmark compared to \work{PromeFuzz} (28 minutes), \work{PromptFuzz} (8 hours), and \work{OGHarn} (1--24 hours). Other tools fuzz each generated harness during synthesis to measure coverage or filter non-functional harnesses, which dominates their inference time. \tool sidesteps this, only verifying that each code block compiles and deferring dynamic exploration to the stitching engine.

Regarding monetary cost, \work{OGHarn} does not use LLMs and so incurs no inference cost. \work{PromptFuzz} with \texttt{gpt-5} was expensive, hitting our 25 USD per-benchmark limit on every benchmark. \tool and \work{PromeFuzz} were cost effective, averaging 1.77 USD and 0.79 USD per benchmark respectively.
Among LLM-based tools, cost per bug was 2.00 USD for \tool, 3.46 USD for \work{PromeFuzz}, and 300.00 USD for \work{PromptFuzz}.

\subsection{RQ2: Real-World Automatic Bug Finding}
\label{sec:rq2}

To evaluate \tool in the wild, we deployed it on 1365 real-world repositories, selecting well-tested and widely-used projects. We avoided small, unmaintained codebases where finding bugs is trivial. 
We selected git repositories consisting of every C/C++ project integrated into
\work{OSS-Fuzz} (only the URLs, not the harnesses), every Chromium dependency, and popular C/C++ libraries.
For popular libraries, we selected the top 400 each of the most-starred C and C++ projects on GitHub
that are actively maintained (received commits in the last month) and self-describe as libraries or frameworks. 
A full list of tested projects is provided in the appendix (\autoref{tab:rq2-list}).

\customsection{Configuration, Inference, and Runtime}
We ran the \agent{Configure Agent} on each repository. It succeeded on 915/1365 (67\%) targets with an average time of 7 minutes and cost of 0.34 USD per project. 
The \textbf{Inference Workflow} ran on the 915 configured projects, identifying an average of 38 objects and 181 functions per target. Inference averaged 15 minutes and 2.39 USD per project, with cost strongly correlated to program size (r=0.901, ~1.3 cents per function).

We then ran the full \emph{stitching} pipeline with \agent{Triage Agent} and \agent{Patch Agent} as applicable, restarting the fuzzer after each patch. Each target was fuzzed on a single vCPU for 1-3 days on a dedicated server with two Intel Xeon Gold 6430 processors and 1 TB RAM.

\customsection{Results}
\tool generated 4556 total reports, marking 1185 as duplicates for 3371 unique crashes: 178 security (p1), 766 stability (p2), and 2427 misuse (p3). The 944 p1/p2 bugs covered 317 (35\%) of targets.
The \agent{Triage Agent} synthesizes fully reproducible testcases and detailed descriptions. All bug reports include a description, minimized testcase, crash report, and dockerized build for easy reproduction.

To respect maintainers' time, rather than automatically submitting these reports via GitHub issues or trackers, we manually reviewed every report before submission. We curated those we were confident were legitimate and security-relevant. We also reviewed each project's security policy to report through appropriate channels before doing so.

To date, we have reviewed ~170 reports and submitted \textbf{131} bug reports across 102 projects (excluding those submitted in RQ1). Of these, \textbf{\numpatched} have been patched and only \textbf{3} marked invalid. A full table is in the appendix (\autoref{tab:rq2-bugs}).

\customsection{Impact}
\tool has discovered serious bugs in important projects:
two out-of-bounds read vulnerabilities in \texttt{opencv}
(one of the most widely used computer-vision libraries);
three bugs in \texttt{libjpeg-turbo} (a popular jpeg library),
including a gap in the documentation that led users to implement
code susceptible to an attacker-controlled heap overflow;
and a \emph{remotely accessible} attacker-controlled heap overflow vulnerability in a release candidate for \texttt{libcups}
(one of the most widely used systems for printer networking).

\section{Related Work}
\label{sec:relwork}

\customsection{Automated Fuzz Harnessing}
There have been many efforts to reduce the manual burden of fuzz harness development through automated techniques. Tools like \work{FUDGE}~\cite{fudge} and \work{FuzzGen}~\cite{fuzzgen} analyze existing consumer code to understand how to invoke a target. \work{UTOPIA}~\cite{utopia} performs a similar type of analysis but starts from unit tests. \work{OGHarn}~\cite{ogharn} relieves the dependency of consumer code through clever heuristic pruning, but is subsequently limited to certain types of easily-fuzzable functions.

\customsection{LLM-Driven Fuzz Harnessing}
With the rapid development of LLMs, it has become easier and easier to simply prompt an LLM to produce a fuzz harness. Recent works focus primarily on the challenge of expanding code coverage and ensuring validity. \work{PromptFuzz}~\cite{promptfuzz} runs harnesses during synthesis using that information to select which functions to use in the next harness prompt, \work{CKGFuzzer}~\cite{ckgfuzzer} introduces a code knowledge graph for better library understanding, and \work{PromeFuzz}~\cite{promefuzz} uses a whole array of techniques including retrieval-augmented generation and an LLM-based bug analyzer. LLM-based fuzz harnessing has also been explored in industry with Google's \work{OSS-Fuzz-Gen}~\cite{oss-fuzz-gen} which has successfully landed new fuzz harnesses in the codebase of real projects.

\customsection{Flexible Fuzzing}
The most directly related works however, are techniques that can flexibly invoke different API sequences at runtime, not bound to pre-selected fixed sequences. \work{Hopper}~\cite{hopper} is one such work, which effectively blindly attempts to run different sequences of functions and heuristically tries to filter out crashes. In practice, though it is often too noisy to be effective at finding bugs (indeed the authors of \work{Hopper} turned to LLMs and developed \work{PromptFuzz} shortly thereafter).
\work{GraphFuzz}~\cite{graphfuzz}, however, is very conceptually similar to \tool, using a specification to guide generation of dataflow graphs, which are analogous to \tool's concept of a ``stitched testcase.'' The core differences are twofold. Firstly, while \work{GraphFuzz} understands input/output object constraints, it has no mechanism to enforce more complicated semantic constraints, which appear ubiquitously in real-world software. Our core contribution here is the development of the extrinsic typestate system, allowing for specification of rich, open-ended constraints. Secondly, building a \work{GraphFuzz} specification is a tedious, manual process. The library does contain a \texttt{doxygen}-based tool to help propose an initial specification, but in most cases it is not even close to plug-and-play. The original paper, for example, evaluated on only five different benchmarks. In contrast, with \tool's LLM-powered infrastructure it can scale fully automatically to literally thousands of diverse targets, without any manual intervention.

\section{Discussion}
\label{sec:discussion}

\customsection{Architecture and Model Capabilities}
\tool's multi-stage architecture (\autoref{fig:llm}) reflects a pragmatic factoring of a complex task into subtasks that current LLMs can handle reliably, each scoped so the model operates with sufficient context and produces mechanically validatable output. This factoring is not fundamental to \emph{stitching} itself; as models grow more capable, we expect much of this scaffolding to simplify or collapse. The stitching formalism and fuzzing engine, by contrast, are model-independent contributions that we expect to remain useful regardless of how the surrounding synthesis pipeline evolves. 

\customsectionnoperiod{What Counts as a Bug?}
A central dilemma in automated bug finding is accurately determining what constitutes a bug. API contracts are expressed in natural language, library users are expected to follow undocumented conventions, and while clearcut cases exist, most live in a context-dependent grey area. \tool encountered a striking example in \texttt{libjpeg-turbo}: the documentation was slightly ambiguous about how a certain flag could be used with an image transformation function, so \tool naturally wrote a code block that sometimes used this flag and found a crash. Upon review, the crash turned out to be an attacker-controlled heap overflow that would affect any user handling untrusted JPEG data with this flag, likely resulting in remote code execution. The maintainers, however, classified it as intended behavior: the flag was simply not allowed in that context, and they patched the documentation to clarify. We searched GitHub and found several codebases that used the library in exactly this way. This case illustrates that bugs can live not just in code but in \emph{documentation}, and that the ``bug-ness'' of a library depends not only on its implementation but on how its users actually call it. It also suggests a natural extension of our work: testing not just libraries in isolation, but validating their usage at call sites in downstream projects.

\customsection{Toward Autonomous Software Maintenance}
\tool currently operates in a loop of \emph{finding} and \emph{reporting} bugs; a natural extension is \emph{fixing} them. In our evaluation, the bug reports we send are received and reviewed by human maintainers, who spend valuable time to understand and fix the bugs. Therefore, we explicitly pre-filtered the reports we sent to include only clearly legitimate, actionable cases, rather than small nits or low-impact issues.
However, in a future where AI agents grow increasingly capable of performing software maintenance tasks, we believe that techniques like \tool could be used in systems to help fully close the loop, running autonomously to find and fix bugs, continuously improving the software.

\cleardoublepage
\appendix
\section*{Ethical Considerations}

This work develops and deploys an automated bug-finding tool on over 1000 open-source projects, discovering and reporting real vulnerabilities. We present a stakeholder-based ethics analysis.

\customsection{Stakeholders}
We identify the following stakeholders: (1)~\emph{open-source maintainers}, who receive bug reports and must allocate time to triage and fix them; (2)~\emph{end users} of the affected software, whose security is impacted by both the existence of the vulnerabilities and the manner in which they are disclosed; (3)~\emph{the broader open-source community}, which benefits from improved software quality but could be harmed by premature or irresponsible disclosure; and (4)~\emph{potential adversaries}, who could exploit vulnerabilities if details were disclosed before patches were available.

\customsection{Responsible Disclosure}
To protect end users and the broader community, all bugs were disclosed through responsible channels. Before submitting any report, we reviewed each project's security policy (e.g., \texttt{SECURITY.md}, dedicated security mailing lists, or coordinated disclosure platforms) and followed its prescribed process. For projects without a formal policy, we used standard issue trackers for non-security bugs and private channels for security-sensitive ones. We did not publicly disclose security-sensitive bugs before maintainers had an opportunity to address them. Bug report links are anonymized in this submission to preserve double-blind review.

\customsection{Respecting Maintainer Time}
Although \tool can generate reports fully automatically, we deliberately chose \emph{not} to submit them automatically. We manually reviewed every report before submission to ensure it described a legitimate, reproducible, and actionable bug. We curated submissions to prioritize bugs with clear security or stability relevance, and omitted reports we judged to be low-impact or ambiguous. Each submitted report included a minimized reproducer, a detailed description, and a fully dockerized build environment for easy reproduction. For projects with heavy issue backlogs, we also contributed patches ourselves to reduce the burden on maintainers.

\customsection{Dual-Use Considerations}
As with any vulnerability-finding tool, \tool could in principle be used to discover vulnerabilities for malicious purposes. We believe the benefits---identifying and fixing real bugs in widely deployed software---substantially outweigh this risk. The tool does not introduce novel exploitation capabilities; it finds bugs that already exist and accelerates the process of fixing them. We plan to release \tool as open-source software to enable the security community to apply it to their own projects, consistent with the long-standing norm in the fuzzing community of openly sharing bug-finding tools (e.g., AFL, LibFuzzer, OSS-Fuzz).

\customsection{No Human Subjects}
This work does not involve human subjects, personal data, or deception. All experiments were conducted on publicly available open-source code.

\cleardoublepage

\section*{Open Science}

All of the source code and artifacts necessary to reproduce the experiments in this paper are available. We have four artifacts:

\customsection{\tool}
The source code for \tool is available at \url{https://github.com/hgarrereyn/STITCH}. This package contains both the stitching-based fuzzing engine and the LLM-powered components, as well as a CLI wrapper to orchestrate the entire process.

\customsection{Evaluation Framework (RQ1)}
Our evaluation framework is available at \url{https://github.com/hgarrereyn/stitch_eval} which contains code to run coverage and bug discovery experiments on the tools we evaluated against both locally and in a Google Cloud Kubernetes cluster. We include Terraform scripts to manage infrastructure provisioning to replicate the same environment.

\customsection{LibAFL-Forkserver}
Other targets in our evaluation were fuzzed using the forkserver mode of \work{LibAFL}~\cite{libafl}. Our implmentation of this fuzzer is based on the \texttt{fuzzbench} configuration and adapted only slightly to mirror the way \work{LibAFL} is configured in \tool. The source code is available at \url{https://github.com/hgarrereyn/libafl-forkserver}.

\customsection{Real-world Harnesses (RQ2)}
All of the configurations and generated harnesses for the large-scale real-world evaluation we did in RQ2 are available at \url{https://github.com/hgarrereyn/stitch_harnesses}.

\cleardoublepage
\bibliographystyle{plain}
\bibliography{references}

\begin{thebibliography}{10}

\bibitem{oss-fuzz}
Abhishek Arya, Oliver Chang, Jonathan Metzman, Kostya Serebryany, and Dongge Liu.
\newblock {OSS-Fuzz}.

\bibitem{fudge}
Domagoj Babi{\'c}, Stefan Bucur, Yaohui Chen, Franjo Ivan{\v{c}}i{\'c}, Tim King, Markus Kusano, Caroline Lemieux, L{\'a}szl{\'o} Szekeres, and Wei Wang.
\newblock Fudge: fuzz driver generation at scale.
\newblock In {\em Proceedings of the 2019 27th ACM Joint Meeting on European Software Engineering Conference and Symposium on the Foundations of Software Engineering}, pages 975--985, 2019.

\bibitem{fewshot}
Tom Brown, Benjamin Mann, Nick Ryder, Melanie Subbiah, Jared~D Kaplan, Prafulla Dhariwal, Arvind Neelakantan, Pranav Shyam, Girish Sastry, Amanda Askell, et~al.
\newblock Language models are few-shot learners.
\newblock {\em Advances in neural information processing systems}, 33:1877--1901, 2020.

\bibitem{hopper}
Peng Chen, Yuxuan Xie, Yunlong Lyu, Yuxiao Wang, and Hao Chen.
\newblock Hopper: Interpretative fuzzing for libraries.
\newblock In {\em Proceedings of the 2023 ACM SIGSAC Conference on Computer and Communications Security}, pages 1600--1614, 2023.

\bibitem{quickcheck}
Koen Claessen and John Hughes.
\newblock Quickcheck: a lightweight tool for random testing of haskell programs.
\newblock In {\em Proceedings of the fifth ACM SIGPLAN international conference on Functional programming}, pages 268--279, 2000.

\bibitem{afl++}
Andrea Fioraldi, Dominik Maier, Heiko Ei{\ss}feldt, and Marc Heuse.
\newblock $\{$AFL++$\}$: Combining incremental steps of fuzzing research.
\newblock In {\em 14th USENIX Workshop on Offensive Technologies (WOOT 20)}, 2020.

\bibitem{libafl}
Andrea Fioraldi, Dominik Maier, Dongjia Zhang, and Davide Balzarotti.
\newblock {LibAFL: A Framework to Build Modular and Reusable Fuzzers}.
\newblock In {\em Proceedings of the 29th ACM conference on Computer and communications security (CCS)}, CCS '22. ACM, November 2022.

\bibitem{evosuite}
Gordon Fraser and Andrea Arcuri.
\newblock Evosuite: automatic test suite generation for object-oriented software.
\newblock In {\em Proceedings of the 19th ACM SIGSOFT symposium and the 13th European conference on Foundations of software engineering}, pages 416--419, 2011.

\bibitem{syzkaller}
Google.
\newblock {Syzkaller: An unsupervised coverage-guided kernel fuzzer}.
\newblock \url{https://github.com/google/syzkaller}, 2017.

\bibitem{honggfuzz}
{Google}.
\newblock {honggfuzz: Security oriented software fuzzer}.
\newblock \url{https://github.com/google/honggfuzz}, 2026.
\newblock GitHub repository, Apache-2.0 license; accessed Jan 7, 2026.

\bibitem{graphfuzz}
Harrison Green and Thanassis Avgerinos.
\newblock Graphfuzz: Library api fuzzing with lifetime-aware dataflow graphs.
\newblock In {\em Proceedings of the 44th International Conference on Software Engineering}, pages 1070--1081, 2022.

\bibitem{fuzzilli2}
Samuel Gro{\ss}, Simon Koch, Lukas Bernhard, Thorsten Holz, and Martin Johns.
\newblock Fuzzilli: Fuzzing for javascript jit compiler vulnerabilities.
\newblock In {\em NDSS}, 2023.

\bibitem{fuzzgen}
Kyriakos Ispoglou, Daniel Austin, Vishwath Mohan, and Mathias Payer.
\newblock $\{$FuzzGen$\}$: Automatic fuzzer generation.
\newblock In {\em 29th USENIX Security Symposium (USENIX Security 20)}, pages 2271--2287, 2020.

\bibitem{utopia}
Bokdeuk Jeong, Joonun Jang, Hayoon Yi, Jiin Moon, Junsik Kim, Intae Jeon, Taesoo Kim, WooChul Shim, and Yong~Ho Hwang.
\newblock Utopia: Automatic generation of fuzz driver using unit tests.
\newblock In {\em 2023 IEEE Symposium on Security and Privacy (SP)}, pages 2676--2692. IEEE, 2023.

\bibitem{li2024havoc}
Ao~Li, Madonna Huang, Caroline Lemieux, and Rohan Padhye.
\newblock The havoc paradox in generator-based fuzzing (registered report).
\newblock In {\em Proceedings of the 3rd ACM International Fuzzing Workshop}, pages 3--12, 2024.

\bibitem{oss-fuzz-gen}
Dongge Liu, Oliver Chang, Jonathan metzman, Martin Sablotny, and Mihai Maruseac.
\newblock {OSS-Fuzz-Gen: Automated Fuzz Target Generation}, May 2024.

\bibitem{promefuzz}
Yuwei Liu, Junquan Deng, Xiangkun Jia, Yanhao Wang, Minghua Wang, Lin Huang, Tao Wei, and Purui Su.
\newblock Promefuzz: A knowledge-driven approach to fuzzing harness generation with large language models.

\bibitem{GoogleFuzzTest}
Google LLC.
\newblock {FuzzTest}: A c++ testing framework for fuzz tests.
\newblock \url{https://github.com/google/fuzztest}, 2025.
\newblock Accessed: 2026-01-12.

\bibitem{libfuzzer}
{LLVM Project}.
\newblock {libFuzzer}.
\newblock \url{https://llvm.org/docs/LibFuzzer.html}.

\bibitem{promptfuzz}
Yunlong Lyu, Yuxuan Xie, Peng Chen, and Hao Chen.
\newblock Prompt fuzzing for fuzz driver generation.
\newblock In {\em Proceedings of the 2024 on ACM SIGSAC Conference on Computer and Communications Security}, pages 3793--3807, 2024.

\bibitem{hypothesis}
David~R MacIver, Zac Hatfield-Dodds, et~al.
\newblock Hypothesis: A new approach to property-based testing.
\newblock {\em Journal of Open Source Software}, 4(43):1891, 2019.

\bibitem{randoop}
Carlos Pacheco and Michael~D Ernst.
\newblock Randoop: feedback-directed random testing for java.
\newblock In {\em Companion to the 22nd ACM SIGPLAN conference on Object-oriented programming systems and applications companion}, pages 815--816, 2007.

\bibitem{jqf}
Rohan Padhye, Caroline Lemieux, and Koushik Sen.
\newblock Jqf: Coverage-guided property-based testing in java.
\newblock In {\em Proceedings of the 28th ACM SIGSOFT International Symposium on Software Testing and Analysis}, pages 398--401, 2019.

\bibitem{catlm}
Nikitha Rao, Kush Jain, Uri Alon, Claire Le~Goues, and Vincent~J Hellendoorn.
\newblock Cat-lm training language models on aligned code and tests.
\newblock In {\em 2023 38th IEEE/ACM International Conference on Automated Software Engineering (ASE)}, pages 409--420. IEEE, 2023.

\bibitem{schloegel2024sok}
Moritz Schloegel, Nils Bars, Nico Schiller, Lukas Bernhard, Tobias Scharnowski, Addison Crump, Arash Ale-Ebrahim, Nicolai Bissantz, Marius Muench, and Thorsten Holz.
\newblock Sok: Prudent evaluation practices for fuzzing.
\newblock In {\em 2024 IEEE Symposium on Security and Privacy (SP)}, pages 1974--1993. IEEE, 2024.

\bibitem{ogharn}
Gabriel Sherman and Stefan Nagy.
\newblock No harness, no problem: Oracle-guided harnessing for auto-generating c api fuzzing harnesses.
\newblock In {\em 2025 IEEE/ACM 47th International Conference on Software Engineering (ICSE)}, pages 775--775. IEEE Computer Society, 2025.

\bibitem{vikram2023can}
Vasudev Vikram, Caroline Lemieux, Joshua Sunshine, and Rohan Padhye.
\newblock Can large language models write good property-based tests?
\newblock {\em arXiv preprint arXiv:2307.04346}, 2023.

\bibitem{ckgfuzzer}
Hanxiang Xu, Wei Ma, Ting Zhou, Yanjie Zhao, Kai Chen, Qiang Hu, Yang Liu, and Haoyu Wang.
\newblock Ckgfuzzer: Llm-based fuzz driver generation enhanced by code knowledge graph.
\newblock In {\em 2025 IEEE/ACM 47th International Conference on Software Engineering: Companion Proceedings (ICSE-Companion)}, pages 243--254. IEEE, 2025.

\bibitem{afl}
Michal Zalewski.
\newblock American fuzzy lop.
\newblock \url{https://lcamtuf.coredump.cx/afl/}.
\newblock Accessed: September 10, 2024.

\end{thebibliography}

\section{Appendix}

\begin{table*}
  \centering
  \footnotesize
  \begin{NiceTabular}[color-inside]{|[start=4]Xcr|[start=4]@{}w{c}{4pt}@{}|[start=4]r|[start=4]r|[start=4]r|[start=4]@{}w{c}{4pt}@{}|[start=4]r|[start=4]r|[start=4]r|[start=4]@{}w{c}{4pt}@{}|[start=3]rr|[start=3]rr|[start=3]}
    & & & & \Block{1-3}{\textbf{Inference Time (m)}} & & & & \Block{1-3}{\textbf{Inference Cost (\$)}} & & & & \Block{1-4}{\textbf{Analysis Cost (\$)}} \\

    & & & & & & & & & & & & \Block{1-2}{\circled{1}} & & \Block{1-2}{\circled{2}} \\

    \cline{13-16}

    \textbf{Benchmark} & \textbf{Commit} & \textbf{API} & & \circled{1} & \circled{2} & \circled{3} & & \circled{1} & \circled{2} & \circled{3} & & \textbf{avg.} & \textbf{tot.} & \textbf{avg.} & \textbf{tot.} \\
  
    \cline{1-3} \cline{5-7} \cline{9-11} \cline{13-16}
    \texttt{c-ares} & \texttt{a8c0917} (1.2) & 136 \CIRCLE& & 13  & 64 & 442&  & 1.48  & 1.61& 25.00 & & 0.22 & 0.22 & 0.00 & 0.00 \\
\texttt{cjson} & \texttt{12c4bf1} (1.4) & 78 \CIRCLE& & 9  & 7 & 1338&  & 0.50  & 0.17& 25.00 & & 0.14 & 0.41 & \na & \na \\
\texttt{lcms} & \texttt{676e803} (1.2) & 358 \CIRCLE& & 44  & 54 & 368&  & 3.92  & 2.27& 25.00 & & 0.21 & 1.49 & 0.00 & 0.03 \\
\texttt{libmagic} & \texttt{0fa2c8c} (1.3) & 18 \CIRCLE& & 4  & 6 & 412&  & 0.12  & 0.06& 25.00 & & 0.22 & 0.22 & \na & \na \\
\texttt{libpcap} & \texttt{e17fe06} (1.2) & 188 \CIRCLE& & 11  & 17 & 190&  & 1.18  & 0.18& 25.00 & & 0.19 & 1.54 & \na & \na \\
\texttt{sqlite3} & \texttt{0df847c} (1.2) & 317 \CIRCLE& & 42  & 74 & 242&  & 4.58  & 5.99& 25.00 & & 0.17 & 0.67 & 0.00 & 0.04 \\
\texttt{zlib} & \texttt{ef24c4c} (1.2) & 82 \CIRCLE& & 19  & 10 & 626&  & 0.80  & 0.21& 25.00 & & 0.21 & 0.84 & 0.00 & 0.00 \\
\cline{1-3} \cline{5-7} \cline{9-11} \cline{13-16}
\texttt{libaom} & \texttt{99fcd81} (1.0) & 56 \CIRCLE& & 7  & 10 & 428&  & 1.41  & 0.20& 25.00 & & 0.27 & 1.89 & 0.00 & 0.02 \\
\texttt{libjpeg-turbo} & \texttt{36ac5b8} (1.0) & 148 \CIRCLE& & 17  & 27 & 906&  & 2.04  & 0.44& 25.00 & & 0.28 & 3.38 & 0.00 & 0.02 \\
\texttt{libpng} & \texttt{738f5e7} (1.0) & 257 \CIRCLE& & 41  & 35 & 421&  & 3.17  & 1.76& 25.00 & & 0.36 & 3.20 & \na & \na \\
\texttt{libtiff} & \texttt{fcd4c86} (1.0) & 200 \CIRCLE& & 21  & 27 & 177&  & 1.43  & 0.87& 25.00 & & 0.42 & 16.53 & 0.00 & 0.00 \\
\texttt{libvpx} & \texttt{9514ab6} (1.0) & 48 \CIRCLE& & 10  & 16 & 323&  & 1.55  & 0.32& 25.00 & & 0.16 & 0.63 & \na & \na \\
\cline{1-3} \cline{5-7} \cline{9-11} \cline{13-16}
\texttt{curl} & \texttt{4c50998} (1.0) & 100 \CIRCLE& & 9  & 14 & \na&  & 1.65  & 0.32& 25.00 & & 0.30 & 0.59 & \na & \na \\
\texttt{liblouis} & \texttt{3d95765} (1.1) & 30 \CIRCLE& & 5  & 22 & \na&  & 0.25  & 0.23& 25.00 & & 0.41 & 2.86 & 0.00 & 0.00 \\
\texttt{exiv2} \textcolor{orange}{\footnotesize \texttt{(C++)}} & \texttt{04e1ea3} (1.0) & 939 \CIRCLE& & 48  & 119 & \na&  & 11.78  & 1.13& 25.00 & & 0.29 & 3.23 & 0.00 & 0.01 \\
\texttt{pugixml} \textcolor{orange}{\footnotesize \texttt{(C++)}} & \texttt{06318b0} (1.1) & 342 \CIRCLE& & 33  & 29 & \na&  & 2.42  & 0.56& 25.00 & & 0.20 & 1.20 & \na & \na \\
\texttt{re2} & \texttt{c84a140} (0.9) & 97 \CIRCLE& & 10  & 9 & \na&  & 0.79  & 0.18& 25.00 & & 0.16 & 0.63 & \na & \na \\
\texttt{tinygltf} \textcolor{orange}{\footnotesize \texttt{(C++)}} & \texttt{a5e653e} (1.0) & 58 \CIRCLE& & 39  & 13 & \na&  & 1.39  & 0.16& 25.00 & & 0.13 & 0.13 & \na & \na \\
\cline{1-3} \cline{5-7} \cline{9-11} \cline{13-16}
\texttt{cgltf} & \texttt{de39988} (2.0) & 38 \CIRCLE& & 8  & \na & \na&  & 0.68  & \na& 25.00 & & 0.31 & 1.85 & \na & \na \\
\texttt{hdf5} & \texttt{0394b03} (1.8) & 102 \LEFTcircle& & 8  & \na & \na&  & 1.16  & \na& 25.00 & & 0.23 & 2.97 & \na & \na \\
\texttt{libical} & \texttt{460b8d7} (1.8) & 280 \LEFTcircle& & 31  & \na & \na&  & 2.53  & \na& 25.00 & & 0.21 & 1.05 & \na & \na \\
\texttt{libucl} & \texttt{51c5e2f} (1.9) & 141 \CIRCLE& & 23  & \na & \na&  & 1.23  & \na& 25.00 & & 0.40 & 14.65 & \na & \na \\
\texttt{openexr} \textcolor{orange}{\footnotesize \texttt{(C++)}} & \texttt{d669510} (1.8) & 87 \LEFTcircle& & 11  & \na & \na&  & 0.58  & \na& 25.00 & & 0.20 & 0.98 & \na & \na \\
\texttt{pcre2} & \texttt{a678783} (1.2) & 78 \CIRCLE& & 14  & \na & \na&  & 0.61  & \na& 25.00 & & 0.17 & 0.99 & \na & \na \\
\cline{1-3} \cline{5-7} \cline{9-11} \cline{13-16}
\texttt{faup} & \texttt{3a26d0a} (2.8) & 45 \LEFTcircle& & 4  & \na & \na&  & 0.43  & \na& 25.00 & & 0.14 & 0.14 & \na & \na \\
\texttt{geos} & \texttt{c8b889b} (1.8) & 140 \CIRCLE& & 21  & \na & \na&  & 1.69  & \na& 25.00 & & 0.15 & 2.32 & \na & \na \\
\texttt{lexbor} & \texttt{6c219fe} (1.8) & 47 \LEFTcircle& & 8  & \na & \na&  & 0.35  & \na& 25.00 & & 0.24 & 1.45 & \na & \na \\
\texttt{libfyaml} & \texttt{1f520e6} (2.3) & 170 \CIRCLE& & 44  & \na & \na&  & 3.99  & \na& 25.00 & & 0.39 & 26.06 & \na & \na \\
\texttt{stormlib} & \texttt{6052223} (1.8) & 88 \CIRCLE& & 16  & \na & \na&  & 0.81  & \na& 25.00 & & 0.41 & 6.60 & \na & \na \\
\cline{1-3} \cline{5-7} \cline{9-11} \cline{13-16}
\texttt{ngiflib} & \texttt{db19270} (2.1) & 7 \CIRCLE& & 7  & 7 & \na&  & 0.17  & 0.07& 25.00 & & \na & \na & 0.00 & 0.01 \\
\texttt{ffjpeg} & \texttt{caade60} (4.1) & 40 \CIRCLE& & 4  & 14 & \na&  & 0.50  & 0.13& 25.00 & & \na & \na & 0.00 & 0.01 \\
\texttt{loguru} \textcolor{orange}{\footnotesize \texttt{(C++)}} & \texttt{4adaa18} (2.8) & 87 \CIRCLE& & 11  & 27 & \na&  & 0.52  & 0.28& 25.00 & & 0.14 & 0.14 & \na & \na \\
\texttt{rapidcsv} \textcolor{orange}{\footnotesize \texttt{(C++)}} & \texttt{083851d} (1.0) & 27 \CIRCLE& & 5  & 15 & \na&  & 0.33  & 0.14& 25.00 & & 0.23 & 0.93 & \na & \na \\
    \cline{1-3} \cline{5-7} \cline{9-11} \cline{13-16}
  \end{NiceTabular}
  \caption[...]{Inference time (minutes), inference cost (USD), and analysis cost (USD) for each tool. \circled{1}: \tool, \circled{2}: \work{PromeFuzz}, \circled{3}: \work{PromptFuzz}. \work{OGHarn} (not shown) takes between 1-24 hours for inference and does not incur LLM costs. \work{OSS-Fuzz} (not shown) relies on manually-written harnesses and does not incur LLM costs. Analysis cost shows the average (per-crash) and the total (per-benchmark) cost of analyzing the bug with either \agent{Triage Agent} or \work{PromeFuzz} bug analyzer.}
  \label{table:results-cost-speed}
\end{table*}

\definecolor{bugyellow}{RGB}{240, 190, 30}
\newcommand{\confirmed}{\textcolor{blue}{\faThumbsUp}}
\newcommand{\reported}{\textcolor{bugyellow}{\faExclamationTriangle}}
\newcommand{\fixed}{\textcolor{green!70!black}{\faCheck}}

\definecolor{rowgray}{HTML}{F0F0F0}
\definecolor{rowwhite}{HTML}{FFFFFF}

\clearpage
\onecolumn
\footnotesize
\begin{longtable}{lllllll}
  \caption{New bugs found by \tool during RQ1 and RQ2. \reported: reported, \confirmed: confirmed, \fixed: fixed.}
  \label{tab:rq2-bugs} \\
  \toprule
  \textbf{Exp.} & \textbf{Project} & \textbf{Type} & \textbf{Category} & \textbf{Status} & \textbf{Link} \\
  \midrule
  \endfirsthead

  \toprule
  \textbf{Exp.} & \textbf{Project} & \textbf{Type} & \textbf{Category} & \textbf{Status} & \textbf{Link} \\
  \midrule
  \endhead

  \bottomrule
  \endfoot

  RQ2 & alanxz/rabbitmq-c & uaf & ownership & \confirmed & \href{https://github.com/alanxz/rabbitmq-c/security/advisories/GHSA-mwcm-w43j-pq75}{GHSA-mwcm-w43j-pq75} \\
RQ2 & ampproject/amphtml & oob read & utf-8 & \reported & \href{https://github.com/ampproject/amphtml/issues/40420}{amphtml\#40420} \\
RQ2 & antirez/smaz & oob write & compression & \reported & \href{https://github.com/antirez/smaz/issues/17}{smaz\#17} \\
\textcolor{orange}{RQ1} & aomedia/aom & oob write & image & \reported & \href{https://aomedia.issues.chromium.org/issues/480978101}{\#480978101} \\
RQ2 & aomediacodec/libavif & oob read & image & \fixed & \href{https://github.com/AOMediaCodec/libavif/issues/2886}{libavif\#2886} \\
RQ2 & assimp/assimp & oob read & parsing & \reported & \emph{by email} \\
RQ2 & aubio/aubio & oob read & logic & \reported & \href{https://github.com/aubio/aubio/issues/435}{aubio\#435} \\
RQ2 & blosc/c-blosc2 & null-deref & compression & \reported & \href{https://github.com/Blosc/c-blosc2/issues/699}{c-blosc2\#699} \\
\end{longtable}

\newcommand{\nogen}{\textcolor{red!70!white}{\faExclamationTriangle}}

\clearpage
\onecolumn
\footnotesize
\rowcolors{2}{white}{gray!10}
\begin{longtable}{lX|rr|rrr|rr|rr}
  \caption{Every project tested in RQ2. Time and cost are in minutes and USD, respectively., respectively. \nogen: \agent{Configure Agent} failed to configure the project. API: number of successfully synthesized code blocks in the resulting specification. For the \agent{Traige Agent} and \agent{Patch Agent}, we report the average (per-crash) and total (per-project) cost of analyzing and/or patching the bug.}
  \label{tab:rq2-list} \\
  \hiderowcolors
  \toprule
  & \multicolumn{2}{c|}{\agent{Configure}} & \multicolumn{3}{c|}{\textbf{Inference}} & \multicolumn{2}{c|}{\agent{Triage}} & \multicolumn{2}{c}{\agent{Patch}} \\
  \textbf{Project} & \textbf{Time} & \textbf{Cost} & \textbf{Time} & \textbf{Cost} & \textbf{API} & \textbf{Agv.} & \textbf{Tot.} & \textbf{Avg.} & \textbf{Tot.} \\
  \midrule
  \showrowcolors
  \endfirsthead

  \hiderowcolors
  \toprule
  & \multicolumn{2}{c|}{\agent{Configure}} & \multicolumn{3}{c|}{\textbf{Inference}} & \multicolumn{2}{c|}{\agent{Triage}} & \multicolumn{2}{c}{\agent{Patch}} \\
  \textbf{Project} & \textbf{Time} & \textbf{Cost} & \textbf{Time} & \textbf{Cost} & \textbf{API} & \textbf{Agv.} & \textbf{Tot.} & \textbf{Avg.} & \textbf{Tot.} \\
  \midrule
  \showrowcolors
  \endhead

  \bottomrule
  \endfoot

  \href{https://github.com/0vercl0k/wtf}{0vercl0k/wtf} & \multicolumn{2}{c|}{\nogen} & & & & & & & \\
\href{https://github.com/0xFA11/MultiplayerNetworkingResources}{0xFA11/MultiplayerNetworkingResources} & 1 & 0.08 & 1 & 0.03 & 2 & - & - & - & - \\
\href{https://github.com/1technophile/OpenMQTTGateway}{1technophile/OpenMQTTGateway} & \multicolumn{2}{c|}{\nogen} & & & & & & & \\
\href{https://github.com/3proxy/3proxy}{3proxy/3proxy} & 2 & 0.20 & 15 & 0.81 & 3 & 0.10 & 0.10 & - & - \\
\href{https://github.com/4paradigm/OpenMLDB}{4paradigm/OpenMLDB} & 3 & 0.51 & 5 & 0.40 & 37 & - & - & - & - \\
\href{https://github.com/504ensicsLabs/LiME}{504ensicsLabs/LiME} & 1 & 0.07 & 1 & 0.04 & 4 & - & - & - & - \\
\href{https://github.com/9fans/plan9port}{9fans/plan9port} & \multicolumn{2}{c|}{\nogen} & & & & & & & \\
\href{https://github.com/AFLplusplus/AFLplusplus}{AFLplusplus/AFLplusplus} & 11 & 0.55 & 13 & 0.89 & 13 & - & - & - & - \\
\bottomrule
\end{longtable}

\end{document}